# Controlled manipulation of oxygen vacancies using nanoscale flexoelectricity

Saikat Das[1,2,*], Bo Wang[3], Ye Cao[4,5], Myung Rae Cho[1,2], Yeong Jae Shin[1,2], Sang Mo Yang[4,6], Lingfei Wang[1,2], Minu Kim[1,2], Sergei V. Kalinin[4,5], Long-Qing Chen[3], and Tae Won Noh[1,2,†]

[1]Center for Correlated Electron Systems, Institute for Basic Science (IBS), Seoul 08826, Republic of Korea.
[2]Department of Physics and Astronomy, Seoul National University (SNU), Seoul 08826, Republic of Korea.
[3]Department of Materials Science and Engineering, The Pennsylvania State University, PA 16802, USA.
[4]Center for Nanophase Materials Sciences, Oak Ridge National Laboratory, Oak Ridge, TN, 37831, USA.
[5]Institute for Functional Imaging of Materials, Oak Ridge National Laboratory, Oak Ridge, TN, 37831, USA.
[6]Department of Physics, Sookmyung Women's University, Seoul 04310, Korea.

**Abstract**

Oxygen vacancies, especially their distribution, are directly coupled to the electromagnetic properties of oxides and related emergent functionalities that have implication in device applications. Here using a homoepitaxial strontium titanate thin film, we demonstrate a controlled manipulation of the oxygen vacancy distribution using the mechanical force from a scanning probe microscope tip. By combining Kelvin probe force microscopy imaging and phase-field simulations, we show that oxygen vacancies can move under a stress-gradient-induced depolarisation field. When tailored, this nanoscale flexoelectric effect enables a controlled spatial modulation. In motion, the scanning probe tip thereby deterministically reconfigures the spatial distribution of vacancies. The ability to locally manipulate oxygen vacancies on-demand provides a tool for the exploration of mesoscale quantum phenomena, and engineering multifunctional oxide devices.

[*] saikat.das87@gmail.com
[†] twnoh@snu.ac.kr

**Introduction**

Oxygen vacancies ($V_O^{\cdot\cdot}$) are elemental point defects in oxides, and they generally function as mobile electron donors. At sufficiently high concentrations, they can disturb the ground state and promote emergent functional phenomena, such as superconductivity[1], ferromagnetism[2,3], metal-to-insulator transition[4], and interface conductivity[5]. In addition, the distribution and dynamics of $V_O^{\cdot\cdot}$ play a central role in many oxide-based energy and memory applications[6,7]. Hence, the ability to manipulate $V_O^{\cdot\cdot}$ provides an opportunity to study the evolution of emergent phenomena and control numerous functionalities, which are essential for developing next generation oxide devices[8,9].

Traditionally, modification of the vacancy concentration has been carried out by annealing oxides at high temperature in reducing/oxidising atmosphere. Recent works have shown that at room temperature, the vacancy concentration can instead be locally changed by employing an electrical bias from a scanning probe microscope (SPM) tip[4,10]. Such local manipulation enables reversible nanoscale control of metal-insulator transitions[4] and the modulation of interface conductivity[11]. However, in addition to the change in the vacancy concentration, the application of bias through an SPM tip is often accompanied by charge injection[12,13] and the formation of protons/hydroxyls[14], which complicates the practical implementation of this approach.

Recently, it has been shown that the mechanical force from the SPM tip can also deplete $V_O^{\cdot\cdot}$ from the contact region[15,16]. However, because this mechanical depletion is less understood, the feasibility of using the force from an SPM tip as an active means of manipulating $V_O^{\cdot\cdot}$ has not been explored. An explanation for the mechanical depletion of $V_O^{\cdot\cdot}$ has been proposed through the so-called piezochemical coupling mechanism[15,16,17], which involves the converse Vegard effect and the flexoelectric effect. The converse Vegard effect accounts for the decrease in vacancy concentration due to force-induced lattice compression. Meanwhile, the flexoelectric effect considers the electromigration of positively charged $V_O^{\cdot\cdot}$ under the stress-gradient-induced flexoelectric field. Notably, by definition, the flexoelectric effect refers to the generation of an electrical polarisation by stress-gradients[18,19]. In this context, Yudin and Tagantsev suggested that the flexoelectric field is not a macroscopic electric field[19]; instead, it should be understood as a pseudo-internal field that polarises a medium in the presence of a stress-gradient[20,21,22,23]. Hence, it remains unclear how this flexoelectric field acts on $V_O^{\cdot\cdot}$.

To be able to mechanically manipulate $V_O^{\cdot\cdot}$, we must therefore thoroughly understand how they respond to the mechanical force from the SPM tip. Subsequently, we have to devise a strategy to employ this force to increase as well as decrease the vacancy concentration in a controlled manner.

Here, we demonstrate a controlled manipulation of $V_O^{\cdot\cdot}$ using the mechanical force from an SPM tip. As a model system, we used a homoepitaxial thin film of $SrTiO_3$, which is an archetypal quantum paraelectric oxide with well-known flexoelectric coefficients[24,25]. Using a Kelvin probe force microscopy (KPFM)-based imaging scheme, we show that besides pushing $V_O^{\cdot\cdot}$ away in the vertical direction, the force promotes the lateral transport of vacancies during the movement of the tip. Using phase-field simulations, we argue that this mechanical redistribution of $V_O^{\cdot\cdot}$ is driven by the depolarisation field associated with the stress-gradient-induced flexoelectric polarisation. The depolarisation field underneath the tip promotes the vertical migration of vacancies, whereas around the contact edge, it traps $V_O^{\cdot\cdot}$ that can move with the tip. Furthermore, we demonstrate that altering the tip geometry can tailor this depolarisation field to preferentially allow the lateral transport of $V_O^{\cdot\cdot}$. This approach enables a controlled spatial modulation of vacancies.

## Results

### Probing oxygen vacancies with Kelvin probe force microscopy

We start by discussing the concept of probing distribution of $V_o^{\cdot\cdot}$ with the KPFM technique (Figs. 1 a-c). KPFM is a surface sensitive technique that measures the contact potential difference (CPD) between the tip and SrTiO$_3$ (STO) surface. The CPD is defined as the offset in the respective vacuum energy levels (in units of V)[26]. However, as schematically illustrated in Fig. 1b, this CPD would change if the STO surface contains $V_o^{\cdot\cdot}$, which can be locally accumulated by an electrical poling. This change in the CPD can originate from the workfunction of STO[27], chemical dipoles (including $V_o^{\cdot\cdot}$-electron pairs), and their orientation[28]. Moreover, the tip bias used during the KPFM measurement has been argued to influence the measured CPD[29]. The contribution of each factor cannot be individually separated, which inhibits a quantitative determination of the vacancy concentration with the KPFM technique. Nonetheless, as shown in Fig. 1c, the KPFM contrast across this $V_o^{\cdot\cdot}$-rich region can be argued to scale proportionally with the vacancy concentration[27]. To establish a proof of concept, in the following we elaborate on the application of KPFM technique to study the diffusion of $V_o^{\cdot\cdot}$ using a 14 and 120-unit cell (uc)-thick STO film (details of the film are provided in the Methods section and in Supplementary Fig. 1).

Figures 1d-e show KPFM images after poling the pristine surface of STO films with a tip bias of -5 V. The images of the 14-uc (120-uc)-thick STO film were taken 20 (16) minutes and 380 (360) minutes after poling. An image contrast is clearly visible across the poled region, which is appearing as a dark rectangular patch. This implies surface charging, either by injected charges, protons/hydroxyls, and/or $V_o^{\cdot\cdot}$ [30,31]. The injected charges and protons/hydroxyls are expected to decay with a timescale, which is independent of the STO thickness[31]. In contrast, because of diffusion or surface reaction that enables the recombination of $V_o^{\cdot\cdot}$ with oxygen from the ambient, the vacancy concentration is expected to decrease with a pronounced thickness-dependent timescale[32]. Figures 1d-e indeed show that the contrast across the poled region diminishes over time, and the decrease is largest in the 14-uc-thick STO film. Therefore, we argue that the surface charging is caused by $V_o^{\cdot\cdot}$, which either undergo diffusion or recombine with oxygen from the ambient.

We can distinguish the dominating mechanism that causes the vacancy concentration to decay over time by analysing the time-dependency of the degree of equilibrium, $S(t)$ that can be calculated from the KPFM images (see Supplementary Note 1). The time-dependency of $S(t)$ describes how the surface equilibrates after the electrical poling. $S(t)$ will exhibit a semi-parabolic (linear) time-dependency if diffusion (surface reaction) is the dominating mechanism [32,27]. Figure 1f plots $S(t)$ as a function of time. Evidently, the time-dependencies of $S(t)$ are semi-parabolic for both films–implying that the diffusion of $V_o^{\cdot\cdot}$ causes the vacancy concentration to decrease over time.

Arguably, during the poling, the applied tip bias perturbs the equilibrium $V_o^{\cdot\cdot}$-distribution of the entire film including the surface, which equilibrates through the diffusion of $V_o^{\cdot\cdot}$. This diffusion occurs both along the out-of-plane and in-plane directions. However, due to the high surface sensitivity of the KPFM technique, the surface-bulk diffusion of $V_o^{\cdot\cdot}$ predominantly affects the time evolution of the KPFM contrast, and thus the time-dependency of $S(t)$. Notably, during this surface-bulk diffusion, the repulsive vacancy-vacancy interaction inhibit $V_o^{\cdot\cdot}$ to migrate independently along the out-of-plane direction[33]. Thus, the time-dependency *of $S(t)$* effectively describes how the perturbed volume under the poled area (in Figs. 1d-e) equilibrates (see Supplementary Note 1 for a detailed discussion). Naturally, this volume would be smaller in the thinner STO film, and thus would equilibrate faster. This explains why the KPFM contrast ($S(t)$) diminishes (grows) more rapidly in the 14-uc thick STO than in the 120-uc thick STO film (Figs. 1d-f).

Following the rationale above, we fit the time evolution of $S(t)$ with Fick's 2$^{nd}$ law of diffusion (solid lines in Fig. 1f). From this fitting, we obtained the diffusion coefficient $D$ = 9.4(3) x 10$^{-19}$ and 3.8(2) x 10$^{-18}$ cm$^2$ s$^{-1}$ for the 14-uc and 120-uc-thick STO film, respectively. These values are well in the range of the bulk value $D_{bulk} \approx 10^{-(17\pm3)}$ cm$^2$ s$^{-1}$ (300K-extrapolated)[34], which validates the conceptual schematic depicted in Fig. 1c. Subsequently, we utilised this correlation between the vacancy concentration and KPFM signal to image the vacancy redistribution under an applied bias and force.

**Oxygen vacancy redistribution under applied bias and force**

In this section using the 120-uc-thick STO film, we compare the response of $V_o^{..}$ to the electrical bias and force from the SPM tip. For this study, we employed a sharp tip with a radius of curvature = 25 nm. As depicted in the KPFM image (Fig. 2a), we scanned the left side of a formerly $V_o^{..}$-enriched region with an increasingly positive tip bias (0-5 V) at a contact force = 0.6 µN. Meanwhile, the right side was scanned with an increasing contact force (0.6-8.5 µN) by holding the tip at the ground potential. In either case, the tip crossed the border between the formerly $V_o^{..}$-enriched and pristine regions. The central part of the KPFM image is uniformly dark, which indicates a uniform vacancy distribution. However, a change in the image contrast, which stems from the $V_o^{..}$-redistribution is visible within both electrically and mechanically scanned areas. Notably, these scans together with acquiring the KPFM image took about 20 minutes. Because of their ultra-slow decay as evident from Fig. 1e, during this time $V_o^{..}$ can be assumed to be kinetically frozen. This implies that the $V_o^{..}$-redistribution is due to the applied bias and force.

To quantify this $V_o^{..}$-redistribution, we constructed the vacancy concentration map from Fig. 2a by defining the normalised vacancy concentration, $\text{NVC} = \frac{(V_b - V)}{(V_b - V_{\min})}$. Here, $V_b$ and $V_{\min}$ represent the baseline and minimum KPFM signals, which were extracted from areas indicated by the black (top-right corner) and white (centre) squares, respectively. The reconstructed NVC map is shown in Fig. 2b. The positive (negative) NVC refers to a higher (lower) vacacny concentraion compared to the intrinsic surface concentration of $V_o^{..}$ (NVC = 0, at the top-right corner).

Profiling the NVC map along lines E and M indicates that both the applied bias and force depleted the formerly $V_o^{..}$-enriched region (Fig. 2c). We find that this depletion can be modelled with a phenomenological Boltzmann sigmoid function in the following form,

$$f(x) = A_2 + \frac{A_1 - A_2}{\left[1 + e^{\frac{x - X_o}{M}}\right]} \quad (1)$$

where $A_1$, $A_2$, $X_o$, and $M$ refer to the initial value, final value, center, and decay rate, respectively. The solid lines in Fig. 2c are fits to this function. Table 1 lists the best-fit parameters. This functional analysis suggests that the applied positive bias and force have a same qualitative effect along the lines E and M, respectively. By comparing the $X_o$ (= $0.5A_1$) values from the fits, we obtain a force-voltage equivalence factor of 0.4 V µN$^{-1}$. The Boltzmann sigmoid fits in Fig. 2c also exhibit an onset. Defining this onset as $0.9A_1$ (indicated by stars), on an ad hoc basis, yields a threshold voltage ($V_{th}$) and force ($F_{th}$) of 1.3 V and 2.7 µN or equivalently 1.1 V, respectively. These threshold values are consistent with the activation barrier potential (= 1-1.4 V) for the electromigration of $V_o^{..}$ in STO[35,36], which further validates the force-voltage equivalence relationship.

Surprisingly, electrical and mechanical scans yield the opposite image contrast across borders in the NVC map. As shown in Fig. 2d, along line $E_L$, NVC monotonically decreases with the increasing bias. However, along line $M_L$, NVC varies non-monotonously: first, it increases as the force increases and then gradually drops; the horizontal line in Fig. 2d marks the background (NVC = 0.1 at 0 V and 0.6 µN). We therefore conclude that the electrical scan depletes the pristine surface outside the left border, while the mechanical scan enriches the pristine surface outside the right border.

The force-induced vacancy-enrichment of the pristine region implies that some of the depleted $V_o^{..}$ moved laterally with the tip across the border during the mechanical scan. Additional experiments elaborate that the mechanical scan does not alter the background within the pristine region, and thus rule out the formation of $V_o^{..}$ and triboelectric charging of the STO surface during the mechanical scan (see Supplementary Note 2). To quantify the lateral motion of $V_o^{..}$, in Fig. 2e we therefore show the NVC with the background (NVC at 0.6 µN) subtracted (i.e., ΔNVC) along lines M and $M_L$. Because KPFM is a surface-sensitive technique, the remnant $V_o^{\cdot\cdot}$ on the surface contribute to the NVC map after $V_o^{\cdot\cdot}$-redistribution. Also, the possibility of recombination with oxygen from the ambient during the lateral motion of $V_o^{\cdot\cdot}$ can be ignored based on the following considerations. The influence of the surface reaction process, which could facilitate this recombination is negligible in our film. Furthermore, the use of a grounded tip during the mechanical scan rules out the bias-induced amplification of this surface reaction process[37]. Hence, for a particular force, the ratio between the net gain (along $M_L$) and the net drop in NVC (along M) represents the fraction of depleted $V_o^{\cdot\cdot}$ that moved laterally with the tip. By comparing the maximum net gain (= 0.1), and corresponding drop (= 0.3) at force = 5.7 µN, we estimated that only approximately 1/3 of the depleted $V_o^{\cdot\cdot}$ moved with the tip along the surface, while the rest migrate into the bulk.

Overall, our main observations are as follows. First, the contact force is bifunctional: it depletes the formerly $V_o^{\cdot\cdot}$-enriched region and simultaneously enriches the pristine surface. Second, the mechanical $V_o^{\cdot\cdot}$-redistribution involves a predominant surface-bulk migration and a relatively weaker lateral motion of $V_o^{\cdot\cdot}$ with the tip.

**Modelling the mechanical redistribution of oxygen vacancies**

To understand the mechanical redistribution of $V_o^{\cdot\cdot}$ we first considered two mechanisms–the converse Vegard effect and the flexoelectric effect[15,16]. Recently, the magnitude of these two effects under an applied force from SPM tip has been compared in a $PbTiO_3$ (PTO) thin film[38]. This study suggests that the converse Vegard effect is much weaker than the flexoelectric effect. Notably, both the PTO and STO have comparable flexoelectric and Vegard coefficients[25,38,39], which determine the relative contributions of these two effects for a given force. Based on these considerations, we thus conclude that the flexoelectric effect predominantly causes the mechanical redistribution of $V_o^{\cdot\cdot}$, and the contribution from the converse Vegard effect is marginal.

For gaining a mechanistic understanding of this flexoelectric effect-driven $V_o^{\cdot\cdot}$-redistribution, we performed phase-field simulations; whereby we incorporated the flexoelectric effect[40] and coupled the time-dependent Ginzburg-Landau and the Nernst-Planck equations[41,42]. The simulation was performed assuming that the STO is paraelectric (see Supplementary Note 3). Unlike in the experiment, the SPM tip was assumed to be static, and following the Hertzian model the contact radius was calculated to be 8 nm for a contact force of 4 µN. Initially, $V_o^{\cdot\cdot}$ were assumed to be homogeneously distributed over the entire STO thickness, instead of being localized on the surface. Despite these oversimplified assumptions, our simulation still provides a qualitative insight into how $V_o^{\cdot\cdot}$ respond to mechanical stimuli. A detailed explanation of our model and a discussion on the redistribution of $V_o^{\cdot\cdot}$ under a positive tip bias are included in Supplementary Notes 4-5.

The stress-gradient from the SPM tip locally polarises STO through the flexoelectric effect[24], as evident from the out-of-plane polarisation vector map in Fig. 3a. Since in the simulation the STO film is assumed to be in the paraelectric phase, the polarisation in Fig. 3a purely stems from the flexoelectric effect. This flexoelectric polarisation reaches a maximum (~ 0.04 C m$^{-2}$) underneath the tip and spatially varies both in magnitude and direction. The resulting polarisation bound charge and the associated depolarisation field accordingly redistribute vacancies around the tip-STO contact region. This redistribution can be visualised from the simulated NVC map in Fig. 3b, which plots the in-plane distribution of $V_o^{\cdot\cdot}$ at the surface. Clearly, the vacancy concentration is decreased (enhanced) underneath the tip (around the contact edge).

An intuitive understanding of the simulated $V_o^{\cdot\cdot}$-redistribution can be gained from the component-resolved distribution of the depolarisation field. As shown in Fig. 3c, the z-component, $E_z^{dep}$, points downward (upward) below the tip (around the contact edge), whereas the x-component, $E_x^{dep}$, exhibits a parentheses-like structure: a node at the contact point and antinodes around the contact edge (Fig. 3d). Effectively, $E_x^{dep}$ points inwards, as indicated by the white arrows in Fig. 3d.

The *y*-component (not shown), $E_y^{dep}$, forms an analogous structure to $E_x^{dep}$ but is rotated by 90° in the *x-y* plane. The decrease in the vacancy concentration underneath the tip can be attributed to the downward $E_z^{dep}$ component, which moves positively charged $V_O^{..}$ from the surface into the bulk. In contrast, the combination of the upward $E_z^{dep}$ and inwardly directed $E_{x,y}^{dep}$ components favours the accumulation of $V_O^{..}$ around the contact edge, yielding a net increase in the vacancy concentration.

The simulation results qualitatively explain the characteristics of the mechanical $V_O^{..}$-redistribution in Fig. 2. During a mechanical scan, a spatially extended and strong downward $E_z^{dep}$ field acts over a larger fraction of $V_O^{..}$ underneath the tip, which results in a dominant surface-bulk migration. In contrast, a small fraction of $V_O^{..}$ becomes effectively trapped within a shallow annular region around the contact edge by the upward $E_z^{dep}$ and inward $E_{x,y}^{dep}$. These trapped vacancies can move laterally with the tip from the $V_O^{..}$-enriched region to the pristine region during the tip's lateral motion. Therefore, the contrasting roles of the depolarisation field underneath the tip and around the contact edge corroborate both the dominant surface-bulk migration and the relatively weaker lateral motion of $V_O^{..}$ with the tip along the surface.

Notably, while scanning, the tip redistributes $V_O^{..}$ regardless whether it moves from left-right or right-left. Thus, the mechanical redistribution of $V_O^{..}$ should be understood as an average response of $V_O^{..}$ to the force applied during the trace and retrace. Moreover, the scanning velocity would influence the $V_O^{..}$-redistribution–longer the tip spends in contact with STO, the larger number of $V_O^{..}$ it would redistribute. To check whether these factors could contribute to the weaker lateral motion of $V_O^{..}$, we performed additional experiments, whereby we applied force only during the trace, and varied the scanning velocity (see Supplementary Figs. 9-11). These experiments also yielded a weaker lateral motion but a stronger surface-bulk migration of $V_O^{..}$ –highlighting the dominating influence of depolarisation field underneath the tip. In the following, we illustrate that the depolarisation field around the tip-STO contact junction can be tailored in favour of the lateral motion of $V_O^{..}$, which enables controllably manipulating the vacancy distribution.

**Controlled spatial modulation of oxygen vacancies**

The basic concept of tailoring the SPM tip-induced depolarisation field can be understood with the aid of Fig. 4a, which compares the simulated surface deformation profiles under a static load of 4 μN using two different tip geometries. The upper panel corresponds to the spherical tip (contact radius = 8 nm) that is used in Fig. 3, and the lower panel corresponds to a flat-ended tip (contact radius = 15 nm). Compared to the spherical one, the flat-ended tip usually imparts a weaker stress on the STO surface underneath. Since the downward $E_z^{dep}$ scales with the stress-gradient underneath the tip, the flat-ended tip induces a very small downward $E_z^{dep}$ (Fig. 4b, upper panel). In contrast, the lateral deformation (indicated by curved arrows in Fig. 4a), which controls the depolarisation field around the contact edge, is alike for both geometries. Consequently, the flat-ended tip induces a depolarisation field distribution around the contact edge (Fig. 4b) similar to that of the spherical tip in Figs. 3c-d. Additionally, an enhanced contact radius enlarges its spatial extent. A selective suppression of $E_z^{dep}$ underneath a flat-ended like tip should significantly reduce the surface-bulk $V_O^{..}$ migration. Meanwhile, the extended depolarisation field around the contact edge should improve the lateral transport of $V_O^{..}$.

To validate our proposition, we performed experiments with a sharp and blunt tip. The estimated radius of curvature of this blunt tip is larger than 200 nm (see Supplementary Fig. 7). Thus, it effectively yields a flat contact junction underneath the tip. We used the 120-uc thick STO film in these experiments. Figures 4c-d show the NVC maps after mechanical scans were performed with these tips at a contact force of 9.5 μN. Notably, we scanned both the left and right boundaries between the $V_O^{..}$-enriched and pristine regions.

To compare the sharp and blunt tip-induced $V_o^{\cdot\cdot}$-redistribution we profiled the NVC maps, as indicated by vertical lines in Figs. 4c-d. Figures 4e-f show the corresponding NVC profiles. The overlapping NVC profiles along lines M1/M4 and M2/M3 demonstrate that the $V_o^{\cdot\cdot}$-redistribution is reproducible for both tip geometries. However, the response of $V_o^{\cdot\cdot}$ to the applied force from these two tips are clearly different. The NVC profiles in Fig. 4e exhibit a maximum drop in NVC of $\Delta_{max}^{dec}$ = -0.75 along lines M2 and M3 but no appreciable increase in NVC ($\Delta_{max}^{inc}$) along lines M1 and M4. This implies that the sharp tip strongly depletes the $V_o^{\cdot\cdot}$-enriched regions but barely enriches the pristine regions. This result is in qualitative agreement with that in Fig. 2e, which shows that the fraction of the depleted $V_o^{\cdot\cdot}$ that laterally move with the tip progressively decreases for applied forces larger than 6 µN. The NVC profiles in Fig. 4f, however, exhibit a maximum drop in NVC of $\Delta_{max}^{dec}$ = -0.25 along lines M2 and M3 and increase by $\Delta_{max}^{inc}$ = +0.2 along lines M1 and M4. This implies that approximately 80% of the depleted $V_o^{\cdot\cdot}$ laterally moved with the blunt tip. The strong reduction (improvement) of $\Delta_{max}^{dec}$ ($\Delta_{max}^{inc}$) thus confirms an active suppression of the out-of-plane migration and a simultaneous enhancement in the lateral transportation of $V_o^{\cdot\cdot}$ during scans with the blunt tip. Overall, the ability to deterministically move $V_o^{\cdot\cdot}$ constitutes the first experimental demonstration of a controlled manipulation of $V_o^{\cdot\cdot}$ in an oxide and the resulting two-dimensional spatial modulation.

**Discussion**

Through a combined experimental and theoretical approach, we demonstrated the flexoelectricity-mediated controlled manipulation of oxygen vacancies by the mechanical force from an SPM tip. A deterministic reconfiguration of spatial vacancy profile provides control over the electron density and related electronic correlation effects. This could enable, using an SPM-based all-in-one platform, the investigation of mesoscale quantum phenomena in oxides[43]. Ultimately, this creates the opportunity for developing mechanically sketched oxide devices, and ambipolar mechanical control of device functionalities such as electroresistance states. The voltage-free operation of the SPM tip would thereby eliminate the possibility of surface charging. At this point, we want to emphasise that flexoelectricity is a universal phenomenon, which can occur in any dielectric[18,44]. However, the flexoelectric coefficients of few oxides, such as $SrTiO_3$ and $BaTiO_3$, are currently known[45]. Thus, our work should motivate the study of flexoelectricity in other oxides.

Broadly speaking, our KPFM-based imaging approach offers a time-efficient way of characterising the activation barrier potential for oxygen vacancy migration at room temperature to complement the conventional Arrhenius analysis[46]. Combined with the feasibility of determining the diffusion coefficient, this technique could thus become an essential metrology tool for oxide-based energy and memory research. Furthermore, our theoretical model that couples the phase-field simulations to the Nernst-Planck equation, can be employed to elucidate how depolarisation fields cause oxygen vacancies, electrons, and holes to redistribute. Therefore, the model can be extended to the study of emergent problems such as the domain wall conductivity, high electrical conductivity of morphotropic phase boundaries, and leakage current in ferroelectric oxides[47,48,49,50].

## Methods

### Thin film Growth

SrTiO$_3$ thin films were homoepitaxially grown on TiO$_2$-terminated Nb:SrTiO$_3$ (0.5% wt. doped) substrates using pulsed laser deposition technique. The growth dynamics and thickness were monitored by in-situ reflection high energy electron diffraction (RHEED) technique. The depositions were performed at 1000°C and using an oxygen partial pressure of 5x10$^{-7}$ torr. After deposition, films were annealed at 800°C for an hour in a 1 torr oxygen atmosphere and subsequently cooled down to room temperature at a cooling rate of 20°C /min.

### Kelvin probe force microscopy

KPFM measurements were carried out using the Asylum Research Cypher SPM at room temperature and under ambient conditions. Pt/Ir-coated metallic tips (NANOSENSORS™ PPP-NCHPt) with a nominal spring constant ≈ 40 N/m were used for electrical/mechanical scans and KPFM imaging. The KPFM measurements were obtained in the non-contact mode using a lift height of 30 nm and the typical scan parameters used are as follows: $V_{ac}$ = 1 V (peak-to-peak), $f_{resonance}$ = 250 kHz, and scan rate = 1 Hz. Before each experiment, the spring constant of the cantilever was accurately determined from force−distance measurements and thermal tuning methods. The contact force during mechanical scans was varied accordingly by controlling the set-point voltage.

### Theoretical modelling

To model the oxygen vacancy redistribution by mechanical force, we performed phase-field simulations by coupling the time-dependent Ginzburg-Landau (TDGL) and Nernst-Planck equations.

$$\frac{\partial \mathbf{P}}{\partial t} = -L \frac{\delta F}{\delta \mathbf{P}} \tag{2}$$

$$\frac{\partial [V_O^{\cdot\cdot}]}{\partial t} = \nabla \left( D_{V_O^{\cdot\cdot}} \nabla [V_O^{\cdot\cdot}] + \mu_{V_O^{\cdot\cdot}} [V_O^{\cdot\cdot}] \nabla \phi \right) \tag{3}$$

In equation (2), **P** is the polarisation vector, $L$ is the kinetic coefficient and $F$ is the total free energy of the system, which includes Landau, electric, gradient and flexoelectric energy contributions. In equation (3), $[V_O^{\cdot\cdot}]$, $D_{V_O^{\cdot\cdot}}$, $\mu_{V_O^{\cdot\cdot}}$, and $\phi$ denote the concentration, diffusion coefficient, and mobility of the oxygen vacancies and the electric potential, respectively. Detailed descriptions of the energy functional F and other relevant simulation parameters are presented in the Supplementary Note 4.

To solve equations (2) and (3), the system was discretised into 100Δ$x$ × 50Δ$y$ × 500Δ$z$ grid points to implement the semi-implicit Fourier method[45]. The 120-uc-thick STO film is simulated to be 480Δ$z$ in thickness, while the substrate and air are each 10Δ$z$ in thickness. The parameters for the total free energy of STO were adopted from work of Y. L. Li et al.[46]. Following the work of R. Moos et al.[47], the initial concentration of oxygen vacancies in STO is assumed to be 3.66×10$^{14}$ cm$^{-3}$ based on our thin film growth conditions. In addition, the diffusion coefficient is assumed to be constant with regard to pressure and calculated as 1.23×10$^{-15}$ cm$^2$ s$^{-1}$ at room temperature[47]. A sufficiently long simulation time is used to ensure that the induced polarisation and oxygen vacancy concentration reach a quasi-steady state. At each time step, the electrostatic and elastostatic equilibrium equations are solved under the electric short-circuit[48] and mechanical mixed boundary conditions[49], respectively.

### Data Availibility

The data that support the findings of this study are available from the corresponding authors upon reasonable request.

**References**

1. Kang, H. J. *et al.* Microscopic annealing process and its impact on superconductivity in T′-structure electron-doped copper oxides. *Nat. Mater.* **6,** 224–229 (2007).

2. Rice, W. D. *et al.* Persistent optically induced magnetism in oxygen-deficient strontium titanate. *Nat. Mater.* **13,** 481–487 (2014).

3. Santander-Syro, A. F. *et al.* Giant spin splitting of the two-dimensional electron gas at the surface of $SrTiO_3$. *Nat. Mater.* **13,** 1085–1090 (2014).

4. Szot, K., Speier, W., Bihlmayer, G. & Waser, R. Switching the electrical resistance of individual dislocations in single-crystalline $SrTiO_3$. *Nat. Mater.* **5,** 312–320 (2006).

5. Chen, Y. Z. *et al.* A high-mobility two-dimensional electron gas at the spinel/perovskite interface of $\gamma$-$Al_2O_3$/$SrTiO_3$. *Nat. Commun.* **4,** 1371 (2013).

6. Adler, S. B. Factors Governing Oxygen Reduction in Solid Oxide Fuel Cell Cathodes. *Chem. Rev.* **104,** 4791–4844 (2004).

7. Sawa, A. Resistive switching in transition metal oxides. *Mater. Today* **11,** 28–36 (2008).

8. Muller, D. A., Nakagawa, N., Ohtomo, A., Grazul, J. L. & Hwang, H. Y. Atomic-scale imaging of nanoengineered oxygen vacancy profiles in $SrTiO_3$. *Nature* **430,** 657–661 (2004).

9. Maier, J. Nanoionics: ion transport and electrochemical storage in confined systems. *Nat. Mater.* **4,** 805–815 (2005).

10. Kumar, A. *et al.* Probing surface and bulk electrochemical processes on the $LaAlO_3$-$SrTiO_3$ interface. *ACS Nano* **6,** 3841–3852 (2012).

11. Bark, C. W. *et al.* Switchable induced polarization in $LaAlO_3$/$SrTiO_3$ heterostructures. *Nano Lett.* **12,** 1765–1771 (2012).

12. Bühlmann, S., Colla, E. & Muralt, P. Polarization reversal due to charge injection in ferroelectric films. *Phys. Rev. B* **72,** 214120 (2005).

13. Xie, Y., Bell, C., Hikita, Y. & Hwang, H. Y. Tuning the electron gas at an oxide heterointerface via free surface charges. *Adv. Mater.* **23,** 1744–1747 (2011).

14. Bi, F. *et al.* 'Water-cycle' mechanism for writing and erasing nanostructures at the $LaAlO_3$/$SrTiO_3$ interface. *Appl. Phys. Lett.* **97,** 173110 (2010).

15. Kim, Y. *et al.* Mechanical control of electroresistive switching. *Nano Lett.* **13,** 4068–4074 (2013).

16. Sharma, P. *et al.* Mechanical tuning of $LaAlO_3$/$SrTiO_3$ interface conductivity. *Nano Lett.* **15,** 3547–3551 (2015).

17. Morozovska, A. N. *et al.* Thermodynamics of electromechanically coupled mixed ionic-electronic conductors: Deformation potential, Vegard strains, and flexoelectric effect. *Phys. Rev. B* **83,** 195313 (2011).

18. S. M. Kogan. Piezoelectric effect during inhomogeneous deformation and acoustic scattering of carriers in crystals. *Sov. Phys. Solid. State* **5,** 2069–2070 (1964).

19. Yudin, P. V & Tagantsev, A. K. Fundamentals of flexoelectricity in solids. *Nanotechnology* **24,** 432001 (2013).

20. Catalan, G. *et al.* Flexoelectric rotation of polarization in ferroelectric thin films. *Nat. Mater.* **10,** 963–967 (2011).

21. Lee, D. *et al.* Giant flexoelectric effect in ferroelectric epitaxial thin films. *Phys. Rev. Lett.* **107,** 057602 (2011).

22. Lu, H. *et al.* Mechanical writing of ferroelectric polarization. *Science.* **336,** 59–61 (2012).

23. Lee, D., Yang, S. M., Yoon, J. & Noh, T. W. Flexoelectric rectification of charge transport in strain-graded dielectrics. *Nano Lett.* **12,** 6436–6440 (2012).

24. Zubko, P., Catalan, G., Buckley, A., Welche, P. R. L. & Scott, J. F. Strain-gradient-induced polarization in $SrTiO_3$ single crystal. *Phys. Rev. Lett.* **99,** 167601 (2007).

25. Stengel, M. Surface control of flexoelectricity. *Phys. Rev. B* **90,** 201112 (2014).

26. Sadewasser, S. & Glatzel, T. (eds). *Kelvin probe force microscopy: measuring and compensating electrostatic forces.* (Springer, 2012).

27. Andrä, M. *et al.* The influence of the local oxygen vacancy concentration on the piezoresponse of strontium titanate thin films. *Nanoscale* **7,** 14351–14357 (2015).

28. Liscio, A., Palermo, V. & Samorì, P. Nanoscale quantitative measurement of the potential of charged nanostructures by electrostatic and kelvin probe force microscopy: unraveling electronic processes in complex materials. *Acc. Chem. Res.* **43,** 541–550 (2010).

29. Nielsen, D. A., Popok, V. N. & Pedersen, K. Modelling and experimental verification of tip-induced polarization in Kelvin probe force microscopy measurements on dielectric surfaces. *J. Appl. Phys.* **118,** 195301 (2015).

30. Kim, Y. *et al.* Ionically-mediated electromechanical hysteresis in transition metal oxides. *ACS Nano* **6,** 7026–7033 (2012).

31. Balke, N. *et al.* Exploring local electrostatic effects with scanning probe microscopy: implications for piezoresponse force microscopy and triboelectricity. *ACS Nano* **8,** 10229–10236 (2014).

32. Bieger, T., Maier, J. & Waser, R. Kinetics of oxygen incorporation in $SrTiO_3$ (Fe-doped): an optical investigation. *Sensors Actuators B Chem.* **7,** 763–768 (1992).

33. Schie, M., Marchewka, A., Müller, T., De Souza, R. a & Waser, R. Molecular dynamics simulations of oxygen vacancy diffusion in $SrTiO_3$. *J. Phys. Condens. Matter* **24,** 485002 (2012).

34. Paladino, A. E. Oxidation kinetics of single-crystal $SrTiO_3$. *J. Am. Ceram. Soc.* **48,** 476–478 (1965).

35. Waser, R. Bulk conductivity and defect chemistry of acceptor-doped strontium titanate in the quenched state. *J. Am. Ceram. Soc.* **74,** 1934–1940 (1991).

36. Shin, C. & Yoo, H. Al-doped $SrTiO_3$: Part II, unusual thermodynamic factor and chemical diffusivity. *Solid State Ionics* **178,** 1089–1094 (2007).

37. Kumar, A., Ciucci, F., Morozovska, A. N., Kalinin, S. V. & Jesse, S. Measuring oxygen reduction/evolution reactions on the nanoscale. *Nat. Chem.* **3,** 707–713 (2011).

38. Morozovska, A. N. *et al.* Flexocoupling impact on size effects of piezoresponse and conductance in mixed-type ferroelectric semiconductors under applied pressure. *Phys. Rev. B* **94,** 174101 (2016).

39. Freedman, D. A., Roundy, D. & Arias, T. A. Elastic effects of vacancies in strontium titanate: Short- and long-range strain fields, elastic dipole tensors, and chemical strain. *Phys. Rev. B* **80,** 064108 (2009).

40. Gu, Y., Hong, Z., Britson, J. & Chen, L.-Q. Nanoscale mechanical switching of ferroelectric polarization via flexoelectricity. *Appl. Phys. Lett.* **106,** 022904 (2015).

41. Cao, Y., Shen, J., Randall, C. & Chen, L.-Q. Effect of ferroelectric polarization on ionic transport and resistance degradation in $BaTiO_3$ by phase-field approach. *J. Am. Ceram. Soc.* **97,** 3568–3575 (2014).

42. Wang, J.-J. *et al.* Defect chemistry and resistance degradation in Fe-doped $SrTiO_3$ single crystal. *Acta Mater.* **108,** 229–240 (2016).

43. Cheng, G. *et al.* Electron pairing without superconductivity. *Nature* **521,** 196–199 (2015).

44. Tagantsev, A. K. Piezoelectricity and flexoelectricity in crystalline dielectrics. *Phys. Rev. B* **34,** 5883–5889 (1986).

45. Zubko, P., Catalan, G. & Tagantsev, A. K. Flexoelectric effect in solids. *Annu. Rev. Mater. Res.* **43,** 387–421 (2013).

46. Arrhenius, S. A. Über die dissociationswärme und den einfluß der temperatur auf den dissociationsgrad der elektrolyte. *Z.Physik. Chem.* **4,** 96–116 (1889).

47. Bednyakov, P. S., Sluka, T., Tagantsev, A. K., Damjanovic, D. & Setter, N. Formation of charged ferroelectric domain walls with controlled periodicity. *Sci. Rep.* **5,** 15819 (2015).

48. Oh, Y. S., Luo, X., Huang, F.-T., Wang, Y. & Cheong, S.-W. Experimental demonstration of hybrid improper ferroelectricity and the presence of abundant charged walls in $(Ca,Sr)_3Ti_2O_7$ crystals. *Nat. Mater.* **14,** 407–413 (2015).

49. Seidel, J. *et al.* Electronic properties of isosymmetric phase Boundaries in highly strained Ca-doped $BiFeO_3$. *Adv. Mater.* **26,** 4376–4380 (2014).

50. Cao, Y., Shen, J., Randall, C. & Chen, L.-Q. Effect of multi-domain structure on ionic transport, electrostatics, and current evolution in $BaTiO_3$ ferroelectric capacitor. *Acta Mater.* **112,** 224–230 (2016).

51. Chen, L.-Q. & Shen, J. Applications of semi-implicit Fourier-spectral method to phase field equations. *Comput. Phys. Commun.* **108,** 147–158 (1998).

52. Li, Y. L. *et al.* Phase transitions and domain structures in strained pseudocubic (100) $SrTiO_3$ thin films. *Phys. Rev. B* **73,** 184112 (2006).

53. Moos, R. & Hardtl, K. H. Defect chemistry of donor-doped and undoped strontium titanate ceramics between 1000° and 1400°C. *J. Am. Ceram. Soc.* **80,** 2549–2562 (1997).

54. Li, Y. L., Hu, S. Y., Liu, Z. K. & Chen, L.-Q. Effect of substrate constraint on the stability and evolution of ferroelectric domain structures in thin films. *Acta Mater.* **50,** 395–411 (2002).

55. Li, Y. L., Hu, S. Y., Liu, Z. K. & Chen, L.-Q. Effect of electrical boundary conditions on ferroelectric domain structures in thin films. *Appl. Phys. Lett.* **81,** 427-429 (2002).

**Acknowledgements**

This work was supported by the Institute for Basic Science in Korea (Grant No. IBS-R009-D1). B.W. and L.Q.C acknowledges the support by the National Science Foundation under the grant number DMR-1410714 and by the Penn State MRSEC, Center for Nanoscale Science, under the award NSF DMR-1420620. Y.C. and S.V.K were supported by the U.S. DOE, Office of Basic Energy Sciences (BES), Materials Sciences and Engineering Division (MSED) under FWP Grant No. ERKCZ07 (Y.C., S.V.K.). A portion of this research was conducted at the Center for Nanophase Materials Sciences, which is a DOE Office of Science User Facility. We would like to thank Prof. Jong-Gul Yoon and Prof. Jin-Seok Chung for discussions. We also acknowledge Dr. Luke Sandilands and John Henry Gruenewald for carefully proofreading the manuscript.

**Author Contributions**

S.D. conceived and planned this project under the direction of T.W.N. S.D. grew the films and performed structural characterisation with assistance from M.K. S.D. carried out KPFM measurements assisted by Y.J.S., S.M.Y., and L.F.W. and performed data analysis. B. W. and Y.C performed theoretical modelling under the direction of L. Q. C and S.V.K. M.R.C performed SEM measurements. S.D. and T.W.N wrote the manuscript with inputs from all authors.

**Competing financial interests**

The authors declare no competing financial interests.

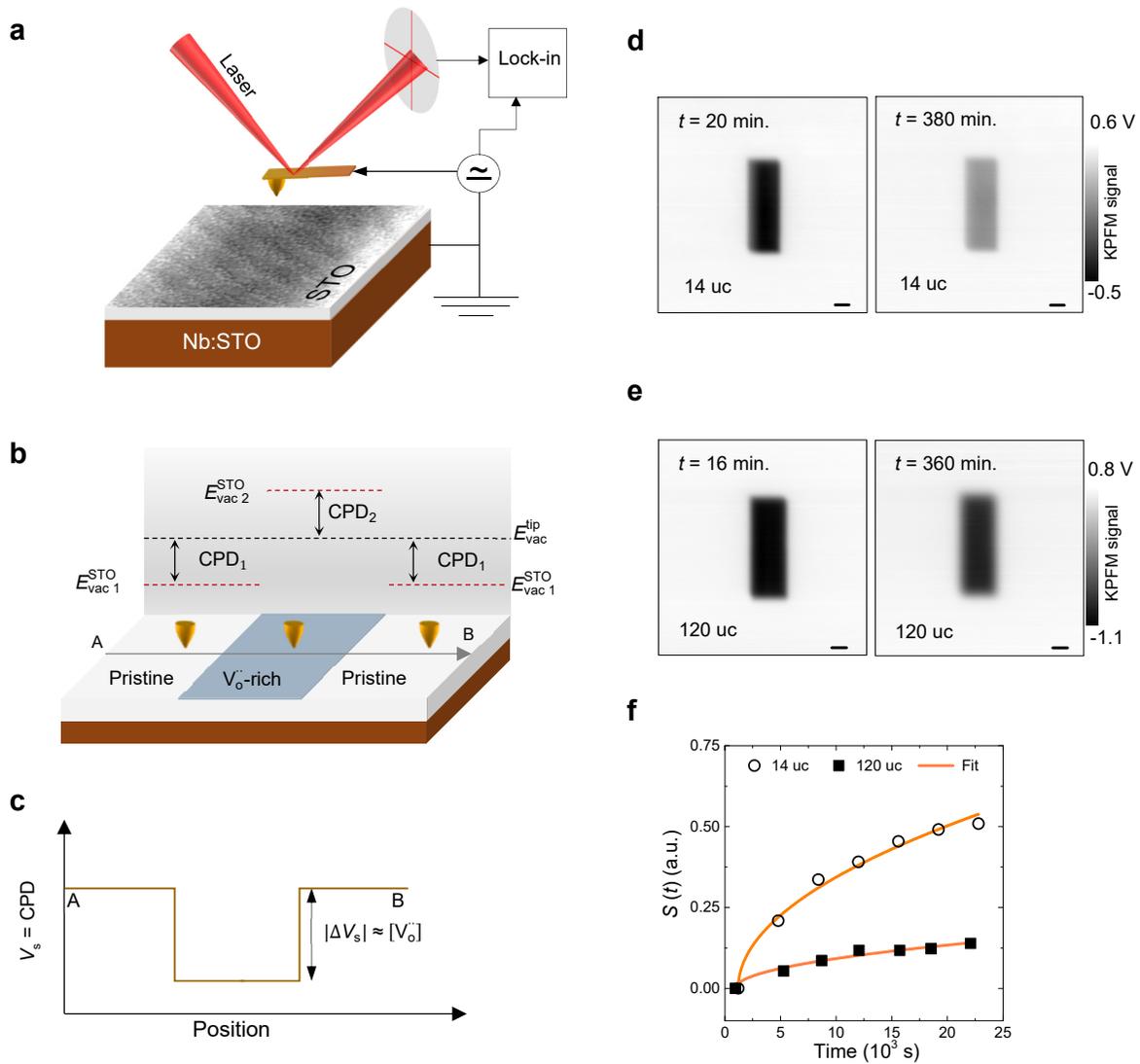

**Figure 1 | Studying diffusion characteristics of oxygen vacancies with KPFM. a,** Sketch of the sample geometry and KPFM measurement architecture. **b,** Schematic illustration of the contact potential difference (CPD) contrast between the pristine (referred to as 1) and $V_o^{..}$-rich (referred to as 2) regions. $E_{vac}$ denotes the vacuum energy level. **c,** Illustrative KPFM signal from a line scan from position A to B across this $V_o^{..}$-rich region. $\Delta V_s$ and $[V_o^{..}]$ denote the net change in the measured KPFM signal and the concentration of $V_o^{..}$ within the $V_o^{..}$-rich region, respectively. **d-f,** Characterisation of diffusion of $V_o^{..}$ with the KPFM technique. KPFM images around a $V_o^{..}$-enriched surface region of a 14-uc-thick (**d**) and 120-uc-thick (**e**) STO films. The time lag between poling the pristine surface and the time of acquiring an image is indicated on the top-right corner of KPFM image. The time evolution of the degree of equilibrium, $S(t)$ (solid symbols) and fit (solid line) according to Fick's 2$^{nd}$ law of diffusion (**f**). The scale bar in **d-e** represents 1 μm.

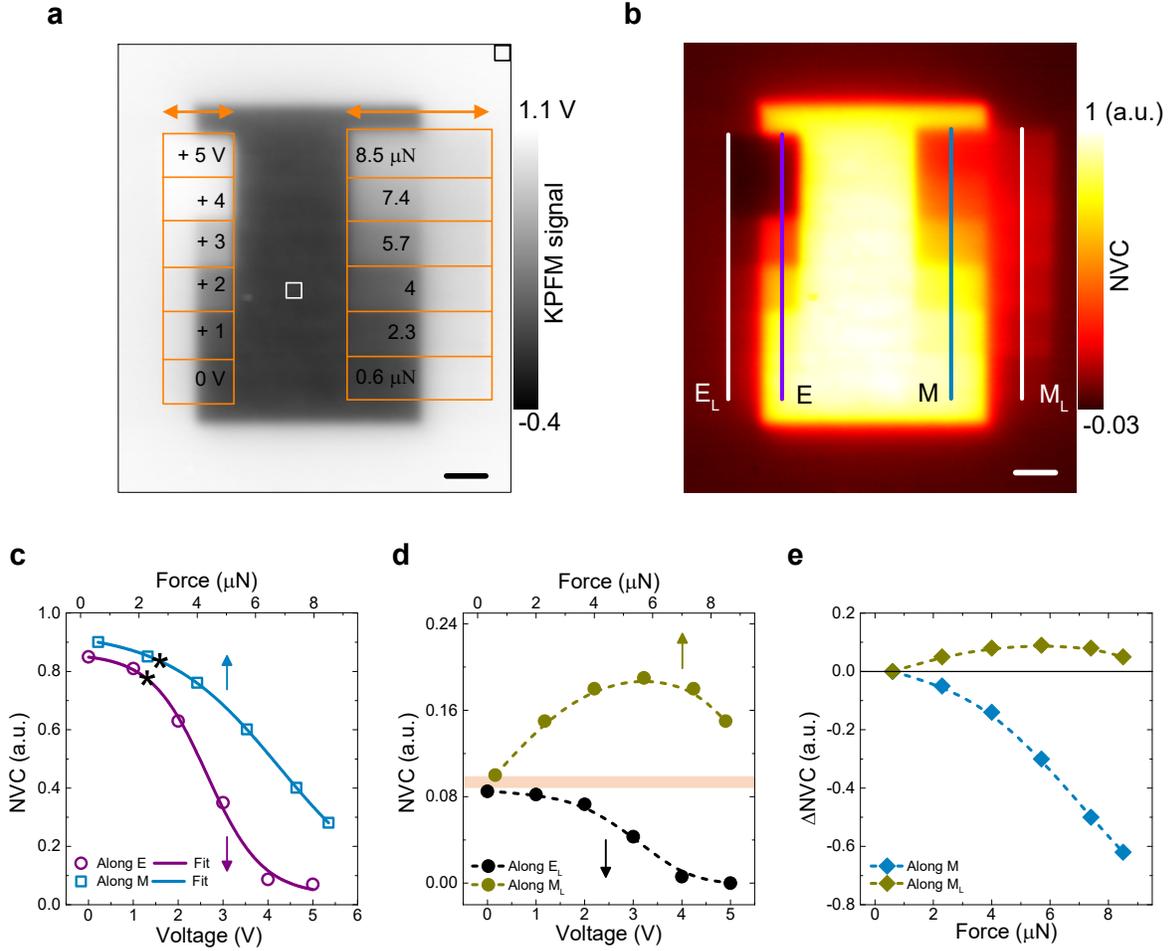

**Figure 2 | Characterisation of oxygen vacancy redistribution by applied bias and force. a,** The KPFM image after electrical and mechanical scans were performed across borders between the $V_o^{\cdot\cdot}$-enriched and pristine regions. The schemes for electrical and mechanical scans are overlaid on the image. Horizontal arrows mark the corresponding fast scan direction. Before electrical and mechanical sans, the $V_o^{\cdot\cdot}$-enrichment was performed by poling a 5x7 µm² area of the pristine surface with a tip bias of -5 V. **b,** The normalised vacancy concentration (NVC) map constructed from the KPFM image in **a**. **c,** The NVC along lines E (open circles) and M (open squares) in **b** measured as a function of applied bias and contact force, respectively. The solid lines denote the Boltzmann sigmoid fit to the data, and Table 1 lists the corresponding best-fit parameters. The threshold voltage ($V_{th}$) and force ($F_{th}$) for the depletion of $V_o^{\cdot\cdot}$ are marked with stars. **d,** The NVC along lines $E_L$ (black coloured circles) and $M_L$ (dark yellow coloured circles) in **b** measured as a function of applied bias and contact force, respectively. The dashed lines are a guide for eyes. Lines $E_L$ and $M_L$ are placed 1 µm away from the borders between the $V_o^{\cdot\cdot}$-enriched and pristine region. The thick horizontal line in **d** marks the background along these lines. **e,** Background-subtracted NVC (ΔNVC) along lines M (turquoise coloured diamonds) and $M_L$ (dark yellow coloured diamonds) in **b**. The dashed lines are a guide for eyes. NVC at 0.6 µN is used as the background. The data plotted in **c** and **d** are extracted using a 0.8 µm-wide averaging window. The scale bar in **a-b** represents 1 µm.

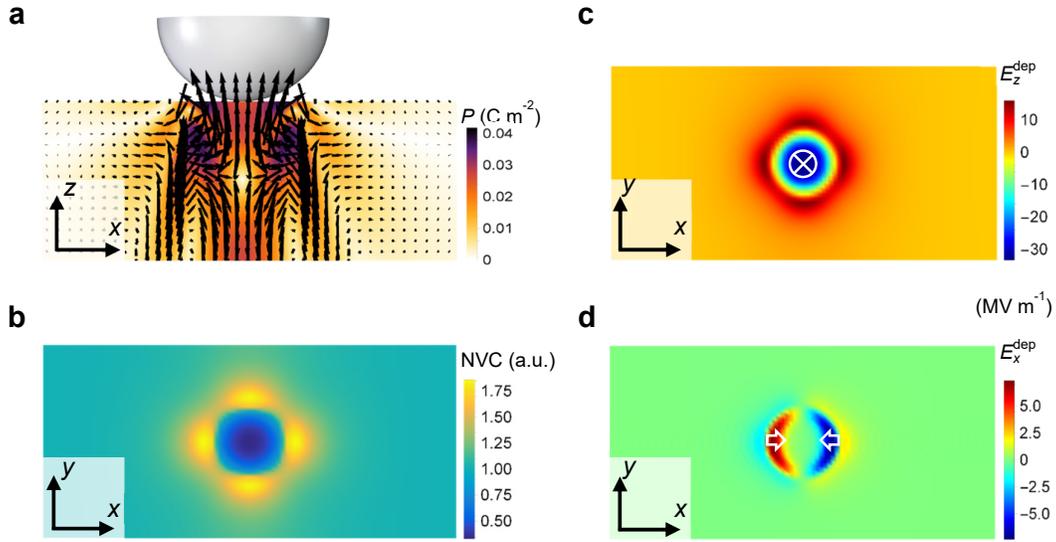

**Figure 3 | Phase-field simulations. a,** Simulated out-of-plane (*z-x* plane) vector map of flexoelectric polarisation induced by the scanning probe microscope (SPM) tip under a static contact force of 4 µN. Arrows denote the direction of the polarisation vectors, and their lengths correspond to the magnitude of the induced polarisation. **b,** Simulated in-plane (*x-y* plane) normalised vacancy concentration (NVC) map around the tip-STO contact region, which shows a depletion and enrichment of $V_o^{\cdot\cdot}$ underneath the tip and around the contact edge, respectively. **c-d,** The component-resolved in-plane distribution of depolarisation field around the tip-STO contact region. The *z*-component, $E_z^{dep}$ (**c**), and the x-component, $E_x^{dep}$ (**d**). The in-plane distribution of the y-component, $E_y^{dep}$, should be viewed as the same as the one in **d** but rotated by 90° in the *x-y* plane.

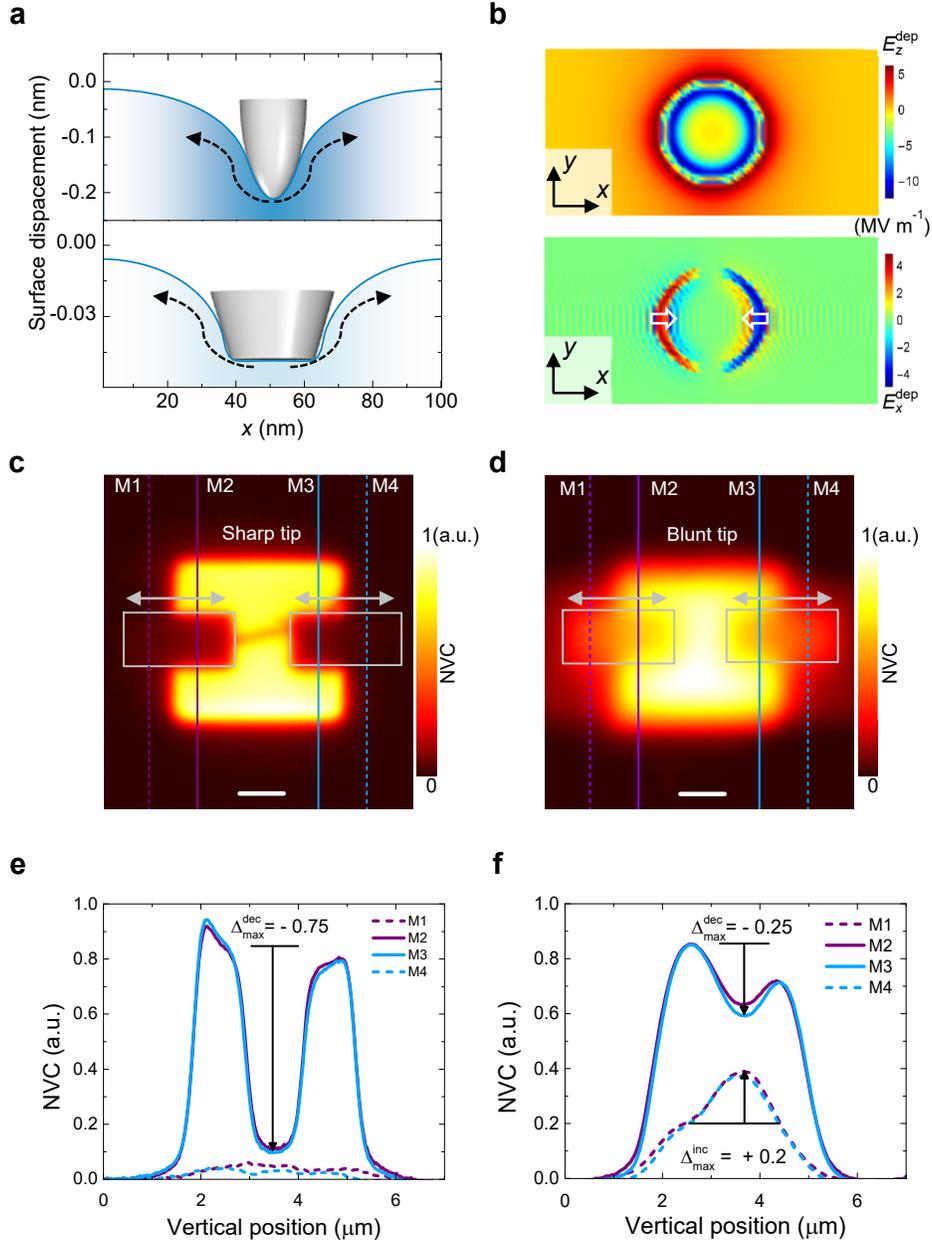

**Figure 4 | Controlled manipulation of oxygen vacancies. a**, Simulated surface deformation profiles under a spherical (upper panel) and flat-ended (lower panel) tip for a static contact force of 4 μN. **b**, Simulated in-plane distribution of the z-component, $E_z^{dep}$ (upper panel) and x-component, $E_x^{dep}$ (lower panel) of the depolarisation field induced by the flat-ended tip. The ripples in $E_x^{dep}$ are numerical artefacts. **c-d**, The normalised vacancy concentration (NVC) maps after mechanical scans were performed using a sharp (**c**) and blunt tip (**d**) with a contact force of 9.5 μN within the grey coloured boxes. Horizontal arrows mark the corresponding fast scan direction. Before mechanical sans, the $V_o^{\cdot\cdot}$-enrichment were performed by poling the pristine surface with a tip bias of -5 V. **e-f**, NVC profiles along lines M1, M2, M3, and M4 in **c**(**e**) and in **d**(**f**). M1 and M4 are placed 0.5 μm away from the borders between the $V_o^{\cdot\cdot}$-enriched and pristine regions. The vertical arrow marks the maximum net increase ($\Delta_{max}^{inc}$) or decrease ($\Delta_{max}^{dec}$) in NVC. Horizontal black lines in **e** and **f** mark the background, which is used to estimate the net change in the NVC. The NVC profiles are averaged over a 0.5-μm-

wide averaging window. Note that the boundaries between the $V_o^{\cdot\cdot}$-enriched and pristine regions in **d** are more diffused compared to those in **c**. This is caused by the use of the blunt tip during the KPFM imaging. The scale bar in **c-d** represents 1 µm.

**Table 1**: **Best-fit parameters from the Boltzmann sigmoid fit to the normalised vacancy concentration (NVC) data in Figure 2c**.

|  | $A_1$ | $A_2$ | $X_0$ | $M$ |
|---|---|---|---|---|
| **Electrical** | 0.86(3) | 0.03(4) | 2.6(1) V | 0.6(1) |
| **Mechanical** | 0.93(1) | 0.004(25) | 6.8(1) µN | 1.9(1) |

**Supplementary Information**

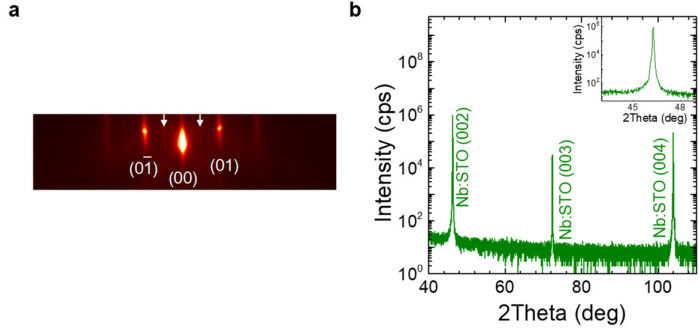

**Supplementary Figure 1 | Structural characterisation of STO film. a,** Reflection high energy electron diffraction (RHEED) pattern along the [100]-azimuth. This image was obtained after the growth of the 120-uc-thick $SrTiO_3$ (STO) film on the Nb:STO substrate. The arrows mark the half-order Bragg reflexes, which typically indicate of a (2x1)-type surface reconstruction. **b,** Symmetric XRD spectrum of the film. The film and substrate peaks are indistinguishable up to the 4$^{th}$-order Bragg Reflex. This confirms stoichiometric film growth. The inset magnifies the XRD spectrum around the (002) peak.

## Supplementary Note 1: Characterisation of $V_o^{\cdot\cdot}$ diffusion using the KPFM technique

Here, adopting a step-by-step approach, we will elaborate on the application of Kelvin probe force microscopy (KPFM) to study the diffusion of oxygen vacancies ($V_o^{\cdot\cdot}$) and determine the corresponding coefficient of diffusion (*D*). First, using the 14-uc-thick $SrTiO_3$ (STO) film as a model system, we discuss how to calculate the degree of equilibrium, $S(t)$, from KPFM images. As we discussed in the main text, the first step of this study was enriching a selected area of the STO surface with $V_o^{\cdot\cdot}$ by poling with a tip bias of -5 V. Then, we performed a series of KPFM measurements over a span of six hours, wherein KPFM images were acquired at regular intervals of approximately one hour. Supplementary Figs. 2a-c show three KPFM images from this series. As time elapsed, the darker, $V_o^{\cdot\cdot}$-enriched region exhibits a change in contrast, which indicates a time-dependent change in the vacancy concentration.

The time evolution of vacancy concentration can be quantified using the degree of equilibrium. This dimensionless quantity can be defined as follows:

$$S(t) = \frac{[V_b(t)-V_{\min}(t)]-[V_b(t_0)-V_{\min}(t_0)]}{[V_b(t_\infty)-V_{\min}(t_\infty)]-[V_b(t_0)-V_{\min}(t_0)]} \quad (1)$$

where $V_b(t)$ and $V_{\min}(t)$ denote the baseline and minimum KPFM signal, respectively. Supplementary Fig. 2d plots horizontal line profiles across the KPFM images in Supplementary Figs. 2a-c. These profiles consist of a plateau and central minimum, which we used to extract $V_b(t)$ and $V_{\min}(t)$, respectively. At any given time, *t*, [$V_b(t)$-$V_{\min}(t)$] corresponds to the concentration of $V_o^{\cdot\cdot}$ within the $V_o^{..}$-enriched region. In Supplementary Equation (1), $t_0$ refers to the time lag between the $V_o^{..}$-enrichment process and the time the first KPFM image was acquired. Meanwhile, $t_\infty$ refers to the time required for the vacancy concertation within the enriched region to equilibrate with the pristine concentration: $[V_b(t_\infty)-V_{min}(t_\infty)] = 0$. Following this approach and by analysing all KPFM images from the series, we calculated $S(t)$ for the 14-uc- and 120-uc-thick STO films, which are shown by symbols in Fig. 1e of the main text. From the definition in Supplementary Equation (1), it is clear that $S(t)$ essentially describes how the surface approaches the equilibrium state after the electrical poling. As we outlined in the main text, the surface equilibrates through diffusion of $V_o^{..}$.

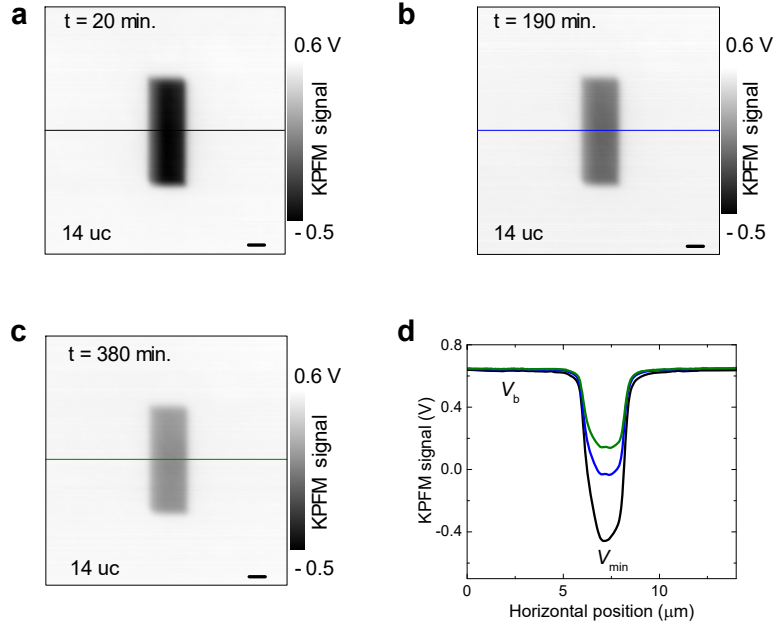

**Supplementary Figure 2 | Time evolution of KPFM signal. a-c,** KPFM images around a $V_o^{..}$-enriched surface region of a 14-uc-thick STO film at 20 minutes (**a**), 190 minutes (**b**), and 380 minutes (**c**) after the $V_o^{..}$-enrichment. The $V_o^{..}$-enrichment was performed by poling the pristine surface with a tip bias of -5 V. **d,** Line profiles from KPFM images in (**a**)-(**c**). Positions used to extract the baseline ($V_b$) and minimum ($V_{min}$) KPFM signal for calculating the equilibrium degree, $S(t)$, are indicated on the graph. The scale bar in **a-c** represents 1 μm.

The $V_o^{..}$ diffuse both along and perpendicular to the film surface. First, to demonstrate the diffusion along the film surface, we show in Supplementary Figs. 3a-b the KPFM images of the 120-uc thick STO film from Fig. 1d of the main text. Supplementary Figs. 3c and e show KPFM profiles along the horizontal and vertical linecuts in Supplementary Figs. 3a-b. These profiles exhibit a broadening that is marked by arrows. For quantifying this broadening, in Supplementary Figs. 3d and f we plot the 1st derivative of these profiles. From a Gaussian fitting (not shown) we found that the full width at half maxima (FWHM) of these derivative profiles increases, in either direction, from about 0.5 to 0.8 μm.

The FWHM of the derivative profile is sensitive to the sharpness of boundaries between the $V_o^{..}$-enriched and pristine regions, and the spatial resolution of the KPFM technique. Since the KPFM images in Supplementary Figs. 3a-b were obtained using identical scanning parameters, and with the same scanning probe microscope (SPM) tip, the spatial resolutions of these two measurements should be identical. Hence, the increase of the FWHM suggests the sharpness of boundaries decreases with time–the $V_o^{\cdot\cdot}$-enriched region laterally expands. This lateral expansion can be attributed to the in-plane diffusion of vacancies.

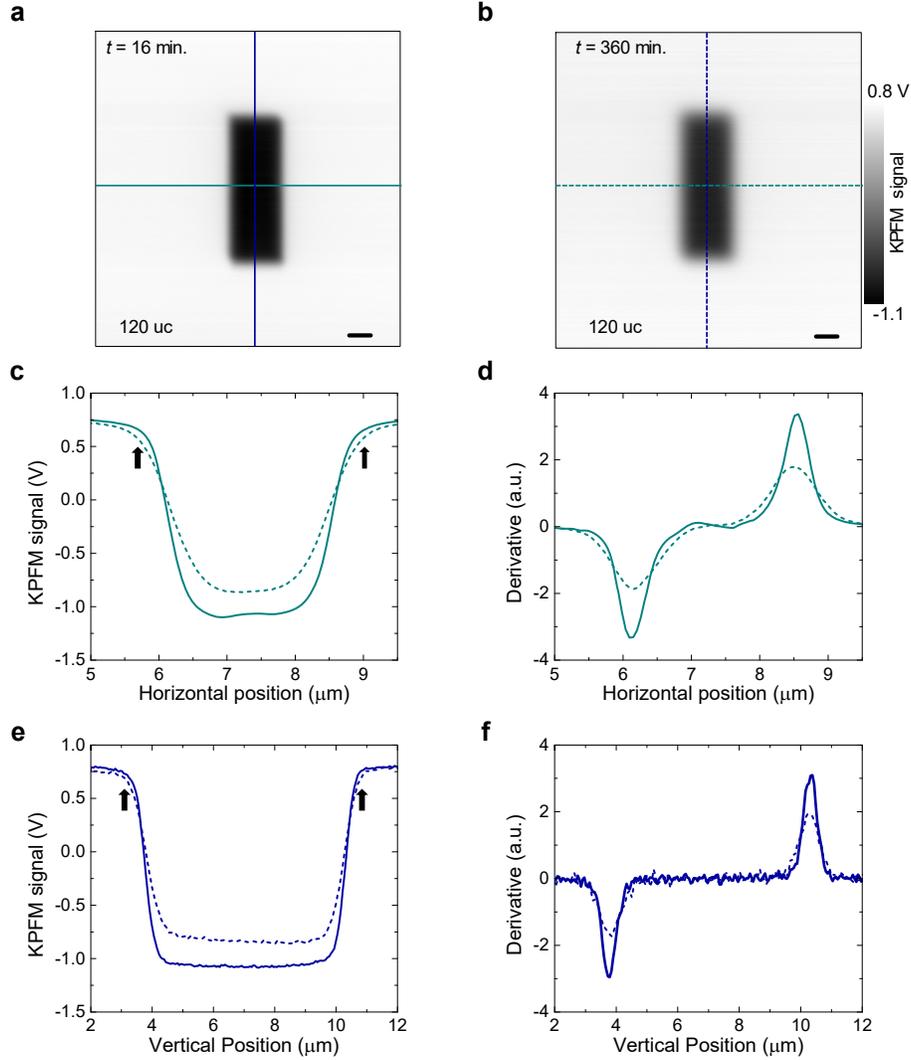

**Supplementary Figure 3 | Evidence of in-plane diffusion of oxygen vacancies. a-b,** KPFM images around a $V_o^{\cdot\cdot}$-enriched surface region of the 120-uc-thick STO films. The time lag between poling the pristine surface and the time of acquiring an image is indicated on the top right corner of KPFM images. Note that **a** and **b** are the same images as those in Fig. 1d of the main text. **c-d,** Horizontal potential profiles (**c**) and their first derivatives (**d**). **e-f,** Vertical potential profiles (**e**) and their first derivatives (**f**). These potential profiles are taken from the KPFM images in **a** and **b**. The arrows in figures (**c**) and (**e**) mark the broadening due to the in-plane diffusion of $V_o^{\cdot\cdot}$. The scale bar in **a-b** represents 1 μm.

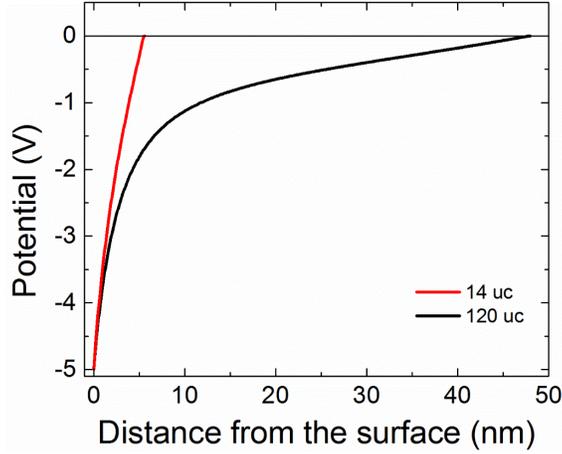

**Supplementary Figure 4 | Simulated depth profiles of electrical potential.** Simulated electrical potential profiles inside the 14-uc and 120-uc thick STO films under an applied tip bias of -5 V. This simulation was performed assuming that the top electrode is biased, while the bottom electrode is grounded, and the relative permittivity of STO is 10. The simulated potential profiles show a gradual decay of electric potential over the thickness.

The KPFM profiles of the 14-uc thick STO film (Supplementary Fig. 2d), however, does not show a discernible broadening. The rapid relaxation of the KPFM contrast inhibits us detecting the broadening, even using the aforementioned FWHM-analysis.

Having elaborated that $V_o^{\cdot\cdot}$ diffuse along the STO surface that is clearly discernible in the thicker STO film; we would like to remark that a meaningful extraction of $D$ from this in-plane diffusion is not possible. The poor spatial resolution of the KPFM technique does not allow us to quantify the lateral expansion of the $V_o^{\cdot\cdot}$-enriched region.

Notably, while calculating $S(t)$ in Supplementary Equation (1), we defined the concentration of $V_o^{\cdot\cdot}$ at any given time, $t$, as $[V_b(t)-V_{min}(t)]$. We used the pristine region and the center of the $V_o^{\cdot\cdot}$-enriched region for extracting $V_b(t)$ and $V_{min}(t)$, respectively. Over time, a certain fraction of $V_o^{\cdot\cdot}$ those were initially located away from boundaries, including those at the center, would certainly diffuse in the lateral directions along the surface. On average this lateral diffusion, however, would not perturb the KPFM signal, and thus the $S(t)$. In contrast, owing to the high surface sensitivity of the KPFM technique, only the surface-bulk diffusion of $V_o^{\cdot\cdot}$ predominantly influences $S(t)$. In the following, we discuss how to extract the corresponding diffusion coefficient ($D$) from $S(t)$.

First, we note that the electrical poling was performed with a tip bias of -5 V, while the bottom electrode (Nb:STO) was grounded. In Supplementary Fig. 4, we compared the depth profiles of the simulated electric potential inside the STO films under the applied tip bias. In this simulation we assumed that the top electrode is biased: $V(x = 0\ nm) = -5$ V and the bottom electrode is grounded: $V(x = $ film thickness in nm$) = 0$ V. Supplementary Fig. 4 shows that the electrical potential gradually decays over the film thickness. The applied tip bias would perturb the equilibrium distribution of $V_o^{\cdot\cdot}$ in STO. Based on the simulated potential profiles shown in Supplementary Fig. 4, we can therefore argue that this perturbation spans the entire film thickness. In this non-equilibrium state, vacancies in larger concentration would accumulate around the surface region. Upon removing the applied bias, the system would relax to its equilibrium state through the vacancy diffusion.

Second, we note a salient feature of $V_o^{..}$ diffusion, which is associated with the electrically charged nature of $V_o^{..}$. Molecular dynamics simulations of $V_o^{..}$-diffusion in STO suggest that during diffusion, the repulsive interaction between charged $V_o^{..}$ inhibits them from moving independently[1].

Summing up above considerations, we can argue that the tip bias applied during the poling perturbs the equilibrium $V_o^{..}$-distribution of the whole STO thickness under the poled area. This perturbed volume equilibrates through diffusion of vacancies, which occurs both along the out-of-plane and in-plane directions. The KPFM technique being surface sensitive, the surface-bulk diffusion of $V_o^{..}$ predominantly causes the KPFM signal or equivalently $S(t)$ to change over time. However, owing to the correlated movement of $V_o^{..}$ during the out-of-plane diffusion of $V_o^{..}$, the time evolution of the KPFM signal or $S(t)$ effectively describes how the perturbed volume under the poled area relaxes to the equilibrium state.

Following the rationale above, we calculated the diffusion coefficient ($D$) by fitting the time evolution of $S(t)$ with Fick's 2$^{nd}$ law of diffusion, which is approximated for small time and one-dimensional cases as follows: $S(t) = \frac{4\sqrt{(t-t_o)D}}{L\sqrt{\pi}}$. Here, $L$ refers to the film thickness, which equilibrates over time through diffusion[2–4]. Note that the use of this one-dimensional approximation is justified because only the out-of-plane diffusion of vacancies influences $S(t)$. Another point worth mentioning concerns the thickness ($L$) dependence of $S(t)$, which for a given $D$ value determines how fast the film equilibrates. The inverse proportionality between $L$ and $S(t)$ justifies the faster temporal evolution of $S(t)$ for the 14-uc thick STO than for the 120-uc thick STO (Fig. 1e in the main text), albeit the calculated $D$ values for these two films are very similar.

**Supplementary Note 2: The effect of mechanical scan on the background in the pristine region.**

To quantify the lateral motion of $V_o^{..}$ along with the tip, in Figs. 2e and 4f of the main text, we used a nonzero normalised vacancy concentration (NVC) in pristine regions along the lines M$_L$, M1, and M4 as our background. Furthermore, we argued that a mechanical scan does not alter this background. In order to support this argument, we performed mechanical scans around two boundaries between a $V_o^{..}$-enriched region and a pristine region of STO surface, as indicated in Supplementary Fig. 5a by boxes A and B. These scans were performed with a sharp tip ($R$ = 25 nm) under a contact force of 5 µN. While box A was scanned, the fast-scan axis was normal to the boundary between the $V_o^{..}$-enriched region and the pristine region. Therefore, the tip crossed this boundary while it moved back and forth. However, while box B was scanned, the fast-scan axis was parallel to the border. Thus, the tip first scanned the $V_o^{..}$-enriched region and then the pristine area. By comparing the resulting change in NVC within the pristine regions enclosed by boxes A and B, we would elaborate that the mechanical scan does not alter the background.

For this comparison, we profiled the NVC map along four sections: lines H1, H2, V1, and V2, as shown in Supplementary Fig. 5a. Supplementary Fig. 5b plots the horizontal NVC profiles H1 and H2, while Supplementary Fig. 5c shows the vertical NVC profiles V1 and V2. The NVC profiles H2 and V2 can be treated as reference profiles, which mark the intrinsic background outside the $V_o^{..}$-enriched part. First, we note that mechanical scans identically deplete the $V_o^{..}$-enriched region within boxes A and B. By profiling (not shown), we estimated the resulting drop in NVC, $\Delta_{max}^{dec}$ > -0.5. However, a direct comparison between Supplementary Figs. 5b and c suggests that the change in NVC is negligible along H1. However, the increase in NVC is sizeable ($\Delta_{max}^{inc}$ = +0.1) along V1, as indicated by a vertical arrow in Supplementary Fig. 5c. These observations validate our argument that the mechanical scan does not alter the background within the pristine region of the NVC map.

The negligible effect of the mechanical scan on the NVC along line H1 (Supplementary Fig. 5b) highlights following points. First and foremost, the mechanical scan does not produce $V_o^{..}$ on the STO surface. Second, the triboelectric charging or contact electrification of the surface due to friction during a mechanical scan is negligible. The presence of strong triboelectric charge would otherwise significantly alter the KPFM signal[5] and hence the NVC along H1. Taking into consideration these points, we can therefore argue that the enrichment of the pristine region by a mechanical scan is caused by $V_o^{..}$ that laterally move with the tip from the $V_o^{..}$-enriched region towards the pristine region.

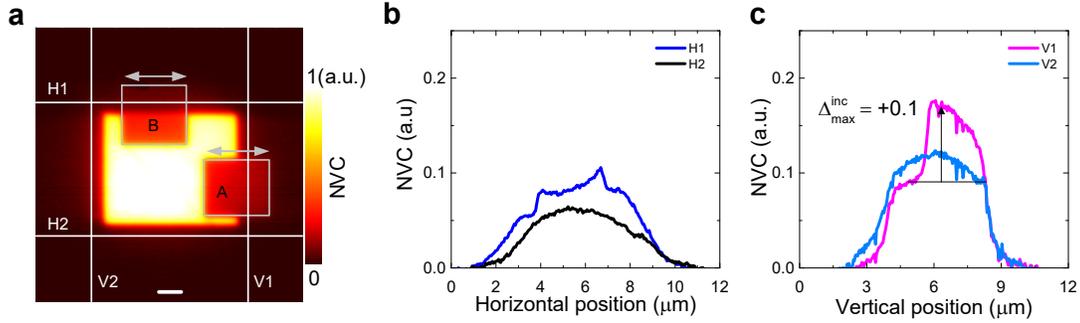

**Supplementary Figure 5 | The scan direction-dependent lateral motion of vacancies. a,** The normalised vacancy concentration (NVC) map after mechanical scans were performed using a sharp SPM tip with a contact force of 5 µN. Scans were performed within areas marked by boxes A and B. Arrows mark the corresponding fast-scan direction. Prior to mechanical scans, the $V_o^{..}$-enrichment was carried out by poling the pristine surface with a tip bias of -5 V. **b,** NVC profiles along the H1 and H2 lines in **a**. **c,** NVC profiles along the V1 and V2 lines in **a**. $\Delta_{max}^{inc}$ refers to the net gain in NVC along the line V1. H1, H2, V1 and V2 lines are placed at a distance 0.5 µm outside the $V_o^{..}$-rich region, and NVC profiles are averaged over a 0.5 µm-wide averaging window. The scale bar in **a** represents 1 µm.

**Supplementary Note 3: PFM characterisation**

In this section, using the piezoresponse force microscopy (PFM) characterisation, we show that our 120-uc-thick STO film is not ferroelectric. First, we note that pure STO is a quantum paraelectric material[6]. However, strain or chemical disorders can drive an STO film into the ferroelectric phase[7,8,9]. Based on our structural characterisation (Supplementary Fig. 1b), we can exclude these extrinsic contributions in our film. Supplementary Fig. 6 shows PFM images that were acquired after obtaining the KFPM image in Fig. 2a of the main text. Vertical (V) and lateral (L) PFM images were obtained using an AC excitation $V_{peak-peak}$ = 0.8 V and a contact resonance frequency in the range of 1.2-1.7 MHz. The uniform $V_o^{..}$-enriched area exhibits a higher PFM amplitude and an 180˚ phase offset relative to the pristine region. On the other hand, changes in the PFM-amplitude and phase reversals are noticeable within the electrically and mechanically scanned parts of the $V_o^{..}$-enriched region.

Remarkably, some features in these PFM images resemble that would be expected from a ferroelectric material, for example, the evolution of the PFM amplitude, particularly within the electrically scanned part. At first, the amplitude decreases, and then it reaches a minimum at applied bias $V_{dc}$ = + 3 V and further increases as the bias increases. This can be clearly observed in Supplementary Fig. 6c, which shows the VPFM amplitude along line E of Supplementary Fig. 6a. Furthermore, the PFM phase also reverses at $V_{dc} \geq 3$ V. Similarly, a VPFM amplitude minimum and an 180˚ phase reversal can be observed along line M (at contact force of 8.5 µN) in the mechanically scanned region. The LPFM images in Supplementary Figs. 6d, e, also contain similar features.

The co-existing amplitude minimum and phase-reversal in a PFM image are hallmarks of ferroelectric domain switching. This domain switching can occur when the electric (flexoelectric) field induced by the tip bias (contact force) exceeds the coercive field[10]. Therefore, one might infer that our STO film is ferroelectric. If this is true, then the ferroelectric polarisation (**P**) must be aligned along the [111] direction, which would then explain the coexisting VPFM and LPFM signals. However, such [111]-oriented **P** has neither been experimentally reported nor theoretically predicted in STO, neither in bulk nor in a thin film[11]. We therefore conclude that the origin of the PFM signal is non-ferroelectric.

To understand the physical origin of PFM signal, we note that for a given phase, *Φ*, the amplitude (*A*) of a PFM signal (= *A cosΦ*) can be expressed as follows:

$$A = A_{\text{piezo}} + A_{\text{el}} + A_{\text{nl}}$$
$$= [d_{ij} + \text{m}(V_{\text{dc}} - V_{\text{cs}}) + \text{n}(V_{\text{dc}} - V_s^{\text{av}})]V_{\text{ac}} \quad (2)$$

In Supplementary Equation (2), the $A_{\text{piezo}}$, $A_{\text{el}}$, and $A_{\text{nl}}$ terms refer to the piezoelectric, local and non-local electrostatic contributions, respectively[12]. For a given AC-excitation, the $A_{\text{piezo}}$ term depends on the material-specific piezoelectric coefficient, $d_{ij}$. In contrast, the electrostatic contribution, $A_{\text{el}}$ ($A_{\text{nl}}$) scales with the offset between the DC-bias, $V_{\text{dc}}$, applied to the probe assembly during a PFM scan and the contact surface potential $V_{\text{cs}}$ (average surface potential observed by the cantilever, $V_s^{\text{av}}$). A non-zero surface potential can arise from surface charges formed during the poling with a biased tip or oxygen vacancy accumulation and the resulting change in the contact potential difference (CPD). Typically, a zero DC-bias ($V_{\text{dc}}$ = 0 V) is used during PFM-imaging. Hence, a non-zero electrostatic contribution is always present in the PFM measurement. For ferroelectric materials with a large piezoelectric coefficient, $A_{\text{piezo}} >> A_{\text{el}}$, $A_{\text{nl}}$, this electrostatic contribution can be neglected. In contrast, while imaging a non-ferroelectric and non-piezoelectric STO film with a zero piezoelectric coefficient ($d_{ij}$), the electrostatic contributions should dominate the PFM signal[13].

Recently, it has been reported that STO films can simultaneously exhibit VPFM and LPFM responses in the presence of oxygen vacancies at the surface[2]. It that report, the PFM response was attributed to the electrostatic contribution due to a non-zero surface potential arising from accumulated vacancies. Our result is consistent with this report. Additionally, by considering the systematic decrease in NVC with applied bias and force from Fig. 2 of the main text, we can correlate the VPFM amplitude with the vacancy concentration. The bias $V_{\text{dc}} \geq 3$ V (contact force of 8.5 μN), for which the VPFM amplitude attains a minimum and the PFM phase reverses, corresponds to a drop in NVC $\geq$ 50%. All these observations therefore point towards a common electrostatic origin of the PFM signal and attest that our STO film is not ferroelectric.

**Supplementary Note 4 : Theoretical modelling of vacancy redistribution under an applied force**

To gain insight into the mechanical redistribution of oxygen vacancies, we numerically solved coupled time-dependent Ginzburg-Landau (TDGL) and Nernst-Planck equations. The time-dependent Ginzburg-Landau equation, which constitutes the basic building block of the phase-field simulation can be expressed as follows:

$$\frac{\partial \boldsymbol{P}}{\partial t} = -L \frac{\delta F}{\delta \boldsymbol{P}} \quad (3)$$

Here, **P** represents the polarisation vector, *L* is the kinetic coefficient, and *F* is the total free energy of the system. The energy functional *F* of the ferroelectric film can be expressed as follows:

$$F = \int (\alpha_{ij} P_i P_j + \alpha_{ijkl} P_i P_j P_k P_l + \frac{1}{2} c_{ijkl} \varepsilon_{ij} \varepsilon_{kl} - q_{ijkl} \varepsilon_{ij} P_k P_l - \frac{1}{2} E_i (\varepsilon_0 \kappa_{ij}^b E_j + P_i) +$$
$$\frac{1}{2} g_{ijkl} \frac{\partial P_i}{\partial x_j} \frac{\partial P_k}{\partial x_l} + \frac{1}{2} f_{ijkl} (\frac{\partial P_k}{\partial x_l} \varepsilon_{ij} - \frac{\partial \varepsilon_{ij}}{\partial x_l} P_k)) \, dV \quad (4)$$

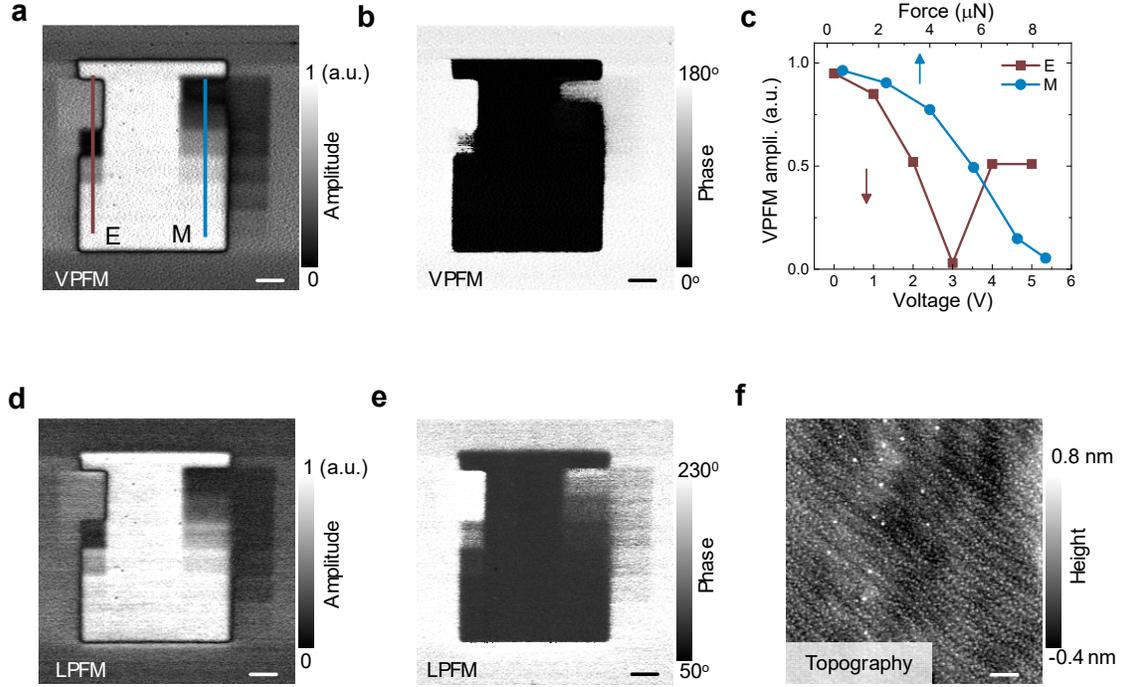

**Supplementary Figure 6 | Piezoresponse force microscopy characterisation. a,** The vertical PFM (VPFM) amplitude image. **b,** VPFM phase image. **c,** The variation of the VPFM amplitude along lines E and M in **a**. **d,** The lateral PFM (LPFM) amplitude image. **e,** LPFM phase image. **f,** Topography image that was acquired in the contact mode during the PFM measurement. These images were subsequently obtained after the KPFM image in Fig. 2a of the main text. No sign of surface deformation due to applied force is present in the topography image **f**. The scale bar in **a**,**b**,**d**,**e**, and **f** represents 1 μm.

Here, $\alpha_{ij}$ and $\alpha_{ijkl}$ denote the second- and fourth-order Landau coefficients, $c_{ijkl}$ is the stiffness tensor, $q_{ijkl}$ is the electrostrictive tensor, $\kappa_{ij}^{b}$ is the background dielectric tensor, $g_{ijkl}$ is the gradient energy coefficient, and $f_{ijkl}$ is the flexocoupling coefficient in units of voltage (V). The numerical values of these coefficients are listed in Supplementary Table 1.

The flexoelectric contribution depends on the flexocoupling coefficients and force-induced strain-gradient (or stress-gradient). Additionally, there is a growing consensus that flexoelectricity could involve a strong contribution from the surface layer[14,15]. Although we did not actively control the surface termination during growth, it is likely that our STO film is $TiO_2$-terminated. We can indirectly support this argument from the RHEED pattern (Supplementary Fig. 1a), which consists of half-order Bragg reflexes that are similar to those obtained for a bare $TiO_2$-terminated Nb:STO substrate surface (images not shown). Furthermore, the atomic force microscopy (AFM) images (not included) of identically grown STO films, with thicknesses ranging from 14 to120-uc, consist of only single-unit cell (0.4 nm) steps. This implies that the STO surface is singly terminated. A mixed $TiO_2$ and SrO termination would otherwise result in half-unit cell steps. These observations led us to use flexocoupling coefficients that are theoretically derived using the first-principle calculation and assuming a $TiO_2$-terminated STO surface[14].

The contact force-induced stress-gradient is calculated by treating the sharp SPM tip as a rigid spherical indenter. The stress imparted by this indenter can be expressed as follows:

$$\sigma_{33}(r) = -\frac{3p}{2\pi a_{\text{sph}}^2}\sqrt{1-\frac{r^2}{a_{\text{sph}}^2}}, \quad r \leq a_{\text{sph}} \tag{5}$$

where $p$ is the applied force, and $a_{\text{sph}}$ is the contact radius. The contact radius $a_{\text{sph}}$ is proportional to $p^{\frac{1}{3}}$, and $r$ is the distance from the contact point.

For the blunt tip, a 12$^{\text{th}}$ order polynomial is used to construct a nearly flat contact surface underneath the tip[16]. The resulting additional surface displacement, $\delta$, can be expressed as follows:

$$\delta(r) = \frac{br^{12}}{12}, \quad r < a_{\text{flat}} \tag{6}$$

where $b$ is a fitting constant, and $a_{\text{flat}}$ is the contact radius of the blunt tip.

The second constituent equation in our model, namely, the Nernst-Planck equation provides insight into the response of $V_O^{\cdot\cdot}$ to flexoelectric polarisation. The Nernst-Planck equation has the following form:

$$\frac{\partial [V_O^{\cdot\cdot}]}{\partial t} = \nabla(D_{V_O^{\cdot\cdot}}\nabla[V_O^{\cdot\cdot}] + \mu_{V_O^{\cdot\cdot}}[V_O^{\cdot\cdot}]\nabla\phi) \tag{7}$$

where $D_{V_O^{\cdot\cdot}}$, $\mu_{V_O^{\cdot\cdot}}$, $[V_O^{\cdot\cdot}]$, and $\phi$ denote the diffusion coefficient, mobility, concentration of oxygen vacancies, and electrical potential, respectively. The diffusion coefficient and mobility are related through the Nernst-Einstein equation, i.e., $D_{V_O^{\cdot\cdot}} = \frac{k_B T}{z_{V_O^{\cdot\cdot}} q_0}\mu_{V_O^{\cdot\cdot}}$, where $z_{V_O^{\cdot\cdot}}$ and $q_0$ represent the valence of oxygen vacancy and elementary electron charge, respectively. The electrical potential $\phi$ is obtained by solving the Poisson equation, $-\nabla^2\phi = \rho_{\text{total}}/\kappa_r^b\varepsilon_0$ under a short-circuit boundary condition at each time step. Here, the total charge density $\rho_{\text{total}}$ consists of both the bound charge associated with the flexoelectric polarisation and the free charge, including electrons, holes and oxygen vacancies.

As a reference, we computed the concentration of electrons, holes and $V_O^{\cdot\cdot}$ from the work of R. Moos et al.[17]. Additionally, we assumed that the concentration of electrons and holes obey the Boltzmann statistics.

$$n = n_0 \exp\left(\frac{\phi q_0}{k_B T}\right) \tag{8}$$

$$p = p_0 \exp\left(\frac{-\phi q_0}{k_B T}\right) \tag{9}$$

In Supplementary Equations (8) and (9), $n_0$ and $p_0$ represent the reference concentration of electrons and holes, respectively. The reference values of the electron, hole, and oxygen vacancy concentration are listed in Supplementary Table 1. Note that all simulated NVC maps in this work are constructed by normalising the concentrations relative to this reference vacancy concentration.

The film surface and the substrate (Nb:STO) were assumed to block oxygen vacancies. Therefore, the Nernst-Planck equation (7) was solved by imposing the boundary condition that the flux ($J_{V_O^{\cdot\cdot}}$) of oxygen vacancies at the film surface ($z = h_2$) and at the film-substrate interface ($z = h_1$) are both zero, i.e.,

$$J_{V_O^{\cdot\cdot}}\big|_{z=h_1,h_2} = -D_{V_O^{\cdot\cdot}}\nabla[V_O^{\cdot\cdot}] - \mu_{V_O^{\cdot\cdot}}[V_O^{\cdot\cdot}]\nabla\phi\big|_{z=h_1,h_2} = 0 \tag{10}$$

**Supplementary Table 1.** Numerical values of parameters used in our theoretical modelling

| Coefficients | Value | Units | Coefficients | Value | Units |
|---|---|---|---|---|---|
| $\alpha_{11}$ | $1.83 \times 10^8$ | $C^{-2}\,m^2\,N$ | $D_{V_O^{\cdot\cdot}}$ | $1.22868 \times 10^{-15}$ | $cm^2\,s^{-1}$ |
| $\alpha_{1111}$ | $1.70 \times 10^9$ | $C^{-4}\,m^6\,N$ | $\mu_{V_O^{\cdot\cdot}}$ | $9.56963 \times 10^{-14}$ | $cm^2\,V^{-1}\,s^{-1}$ |
| $\alpha_{1122}$ | $1.37 \times 10^9$ | $C^{-4}\,m^6\,N$ | $n_0$ | $9.12187 \times 10^{14}$ | $cm^{-3}$ |
| $c_{11}$ | $3.36 \times 10^{11}$ | $N\,m^{-2}$ | $p_0$ | $1.79036 \times 10^{14}$ | $cm^{-3}$ |
| $c_{12}$ | $1.07 \times 10^{11}$ | $N\,m^{-2}$ | $[V_O^{\cdot\cdot}]$ | $3.66576 \times 10^{14}$ | $cm^{-3}$ |
| $c_{44}$ | $1.27 \times 10^{11}$ | $N\,m^{-2}$ | $p$ | 4 | $\mu N$ |
| $q_{11}$ | 0.04581 | $C^{-2}\,m^4$ | $a_{\text{sph}}$ | 8 | nm |
| $q_{12}$ | -0.0135 | $C^{-2}\,m^4$ | $a_{\text{flat}}$ | 15 | nm |
| $q_{44}$ | 0.0096 | $C^{-2}\,m^4$ | $V_{\text{bias}}$ | 1 | V |
| $g_{11}$ | $1.0 \times 10^{-11}$ | $J\,m^3\,C^{-2}$ | $f_{11}$ | 0.8 | V |
| $g_{12}$ | $0.5 \times 10^{-11}$ | $J\,m^3\,C^{-2}$ | $f_{12}$ | -2.63 | V |
| $g_{44}$ | $0.5 \times 10^{-11}$ | $J\,m^3\,C^{-2}$ | $f_{44}$ | -2.01 | V |
| $\kappa_r^b$ | 10 | 1 | | | |

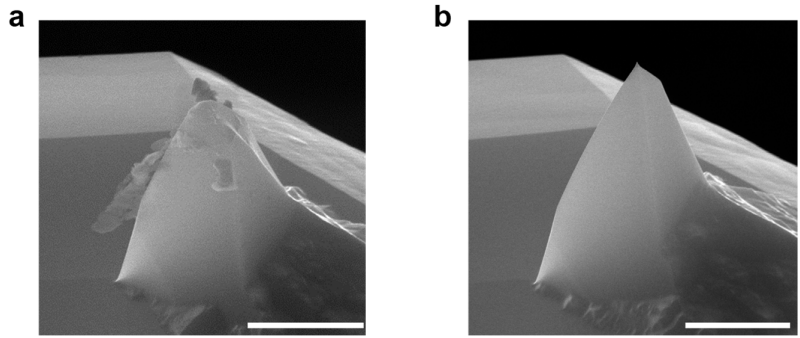

**Supplementary Figure 7 | SEM images of SPM tips. a,** Scanning electron microscope (SEM) image of the blunt tip that was used for the controlled manipulation of oxygen vacancies. We estimated the radius of curvature of this tip to be larger than 200 nm. **b,** The SEM image of an as-received SPM tip, with an estimated radius of curvature = 25 nm. The scale bar in **a-b** represents 5 µm.

**Supplementary Note 5 : Simulation of the redistribution of oxygen vacancies under a positive tip bias**

Our experiment indicates that $V_o^{\cdot\cdot}$ migrate from the surface to bulk in response to a positive bias from the SPM tip. However, in contrast to the mechanical case, $V_o^{\cdot\cdot}$ do not move laterally with the tip (Fig. 2 in the main text). To understand this behaviour of vacancies, we performed phase-field simulations under a positive tip bias. The surface of STO was assumed to be traction-free, and the electrical potential, $\phi$, at the surface was approximated by the following Lorentzian distribution:

$$\phi_{\text{surf}}(r) = \phi_0 \left(\frac{\gamma^2}{\gamma^2 + r^2}\right) \quad (11)$$

Here, $\phi_0$ represents the applied bias voltage on the tip, $\gamma$ is the half-width at half-maximum of the applied bias, and $r$ is the distance from the contact point. Only the bound charge associated with the electric field-induced polarisation was considered when solving the Poisson equation to obtain the electrical potential distribution. An identical boundary condition, as given by Supplementary Equation (10), was used to solve the Nernst-Planck equation.

With an applied bias $\phi_0 = 1$ V and $\gamma = 10$ nm, the computed distributions of out-of-plane ($E_z^{\text{bias}}$) and in-plane ($E_x^{\text{bias}}$) components of the electric field on the film surface are plotted in Supplementary Figs. 8a and b, respectively. The resulting redistribution of vacancies can be visualised from the NVC map in Supplementary Fig. 8c. Because the $E_z^{\text{bias}}$ component acting underneath the tip is directed downward, it depletes $V_o^{\cdot\cdot}$ from the contact junction within a circular region with a radius of approximately 20 nm. In contrast, though the in-plane electrical field may also drive the oxygen vacancies toward lateral directions, our simulation does not show any accumulation of $V_o^{\cdot\cdot}$ on the surface. This can be understood by noting that $E_x^{\text{bias}}$ field is an order of magnitude smaller than $E_z^{\text{bias}}$. Additionally, unlike the force-induced case shown in Fig. 3 of the main text, there is no inwardly (upwardly) directed in-plane (out-of-plane) electric field component around the contact edge that could trap vacancies on the surface. Our simulation therefore qualitatively accounts for the observed out-of-plane migration and the absence of in-plane motion of oxygen vacancies under an applied positive bias.

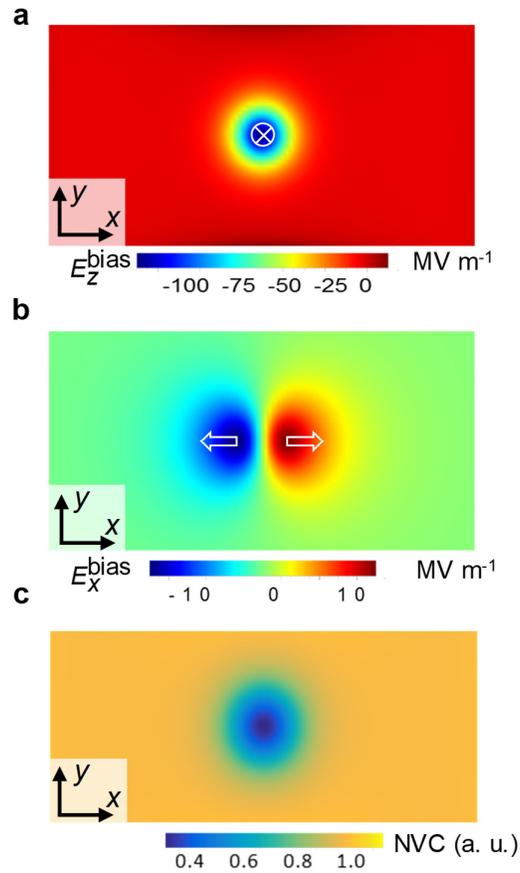

**Supplementary Figure 8 | Simulated electric field and vacancy redistribution under a positive tip bias. a-b,** Simulated electric field under a tip bias of +1 V. The in-plane distribution of the z-component, $E_z^{bias}$ (**a**) and the x-component, $E_x^{bias}$ (**b**). The y-component of the electric field has a similar distribution to $E_x^{bias}$ but is rotated by 90° in the x-y plane. **c,** Simulated NVC map under this electric field. The NVC map shows a strong depletion underneath the tip and no lateral accumulation on the surface.

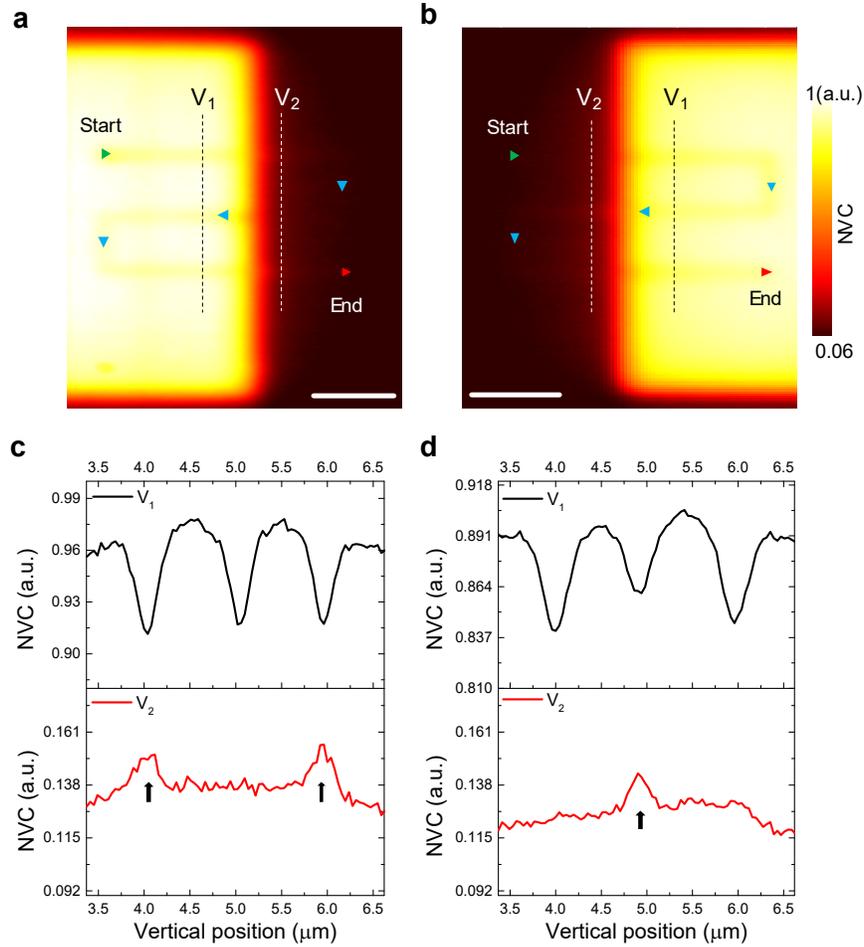

**Supplementary Figure 9 | Oxygen vacancy redistribution by a single line scan. a-b,** The normalised vacancy concentration (NVC) maps after mechanical scans were performed with a load of 6 μN across the left (**a**) right (**b**) boundaries between a vacancy enriched and pristine regions. We used the 120-uc thick STO film and a sharp tip for this experiment. During these scans, we traced the tip following a predefined defined path, which is marked by small triangles. The start (end) of the trace is marked by the green (red) triangle. **c-d,** NVC profiles along lines $V_1$ and $V_2$ in the NVC map in **a** (**c**) and **b** (**d**). The NVC profile along the line $V_1$ ($V_2$) shows a decrease (increase) of $V_O^{..}$ concentration. The vertical arrows in the lower panel of **c** and **d** mark the increase in the vacancy concentration in the pristine regions. This study demonstrates the feasibility of laterally moving vacancies with a single trace. Furthermore, this study elaborates that irrespective of the scan direction, the tip laterally moves $V_O^{..}$ from the $V_O^{..}$-enriched to the pristine region and depletes the former region. Note that the high background inhibits us detecting the lateral motion of vacancies within the $V_O^{..}$-enriched region. Another notable feature concerns the tip's trace from the pristine towards $V_O^{..}$-enriched region, which does not enrich the pristine region. This implies that the mechanical scan does not create $V_O^{..}$ on the STO surface, and also rules out any triboelectric charging of the surface. These arguments are in line with those we discussed in Supplementary Note 2. The scale bar in **a** represents 1 μm.

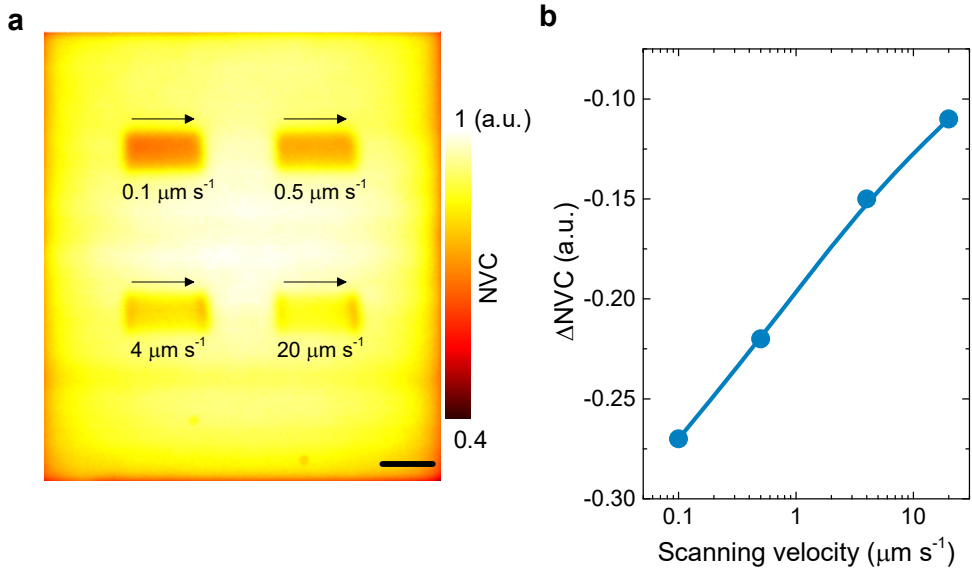

**Supplementary Figure 10 | The scanning velocity dependence of vacancy depletion.**
**a**, The normalised vacancy concentration (NVC) map after mechanical scans were performed with different scanning velocities (0.1-20 µm s$^{-1}$) within a $V_o^{\cdot\cdot}$-enriched region. We used the 120-uc thick STO film and a sharp tip for this experiment. Scans were performed over a 1x0.5 µm$^2$ area with a contact force of 6 µN. The force was applied only during the trace, and the tip was lift-off during the retrace. Fives lines were traced from left-right as marked by horizontal arrows. **b**, Plot of the background subtracted NVC (ΔNVC) as a function of scanning velocity. The solid line is a guide for eyes. Notably, the high background within the $V_o^{\cdot\cdot}$-enriched region inhibits us detecting the lateral motion of $V_o^{\cdot\cdot}$ with the tip. Thus, this experiment only allows us characterising the surface-bulk migration $V_o^{\cdot\cdot}$ as a function of scanning velocity. Figure **b** shows that the net drop in vacancy concentration monotonically increases with decreasing scanning velocity. This implies the longer the tip stays in contact with the $V_o^{\cdot\cdot}$-enriched surface, the larger number of vacancies migrate into the bulk. Note an increased depletion of vacancies around the left and right sides, which is particularly discernible for rectangles scanned with velocities of 4 µm s$^{-1}$ and 20 µm s$^{-1}$. We believe that this larger depletion is due to a stronger deformation of the STO surface during the tip's engagement and withdrawal from the surface. We have carefully excluded these regions while plotting ΔNVC in **b**. The scale bar in **a** represents 1 µm.

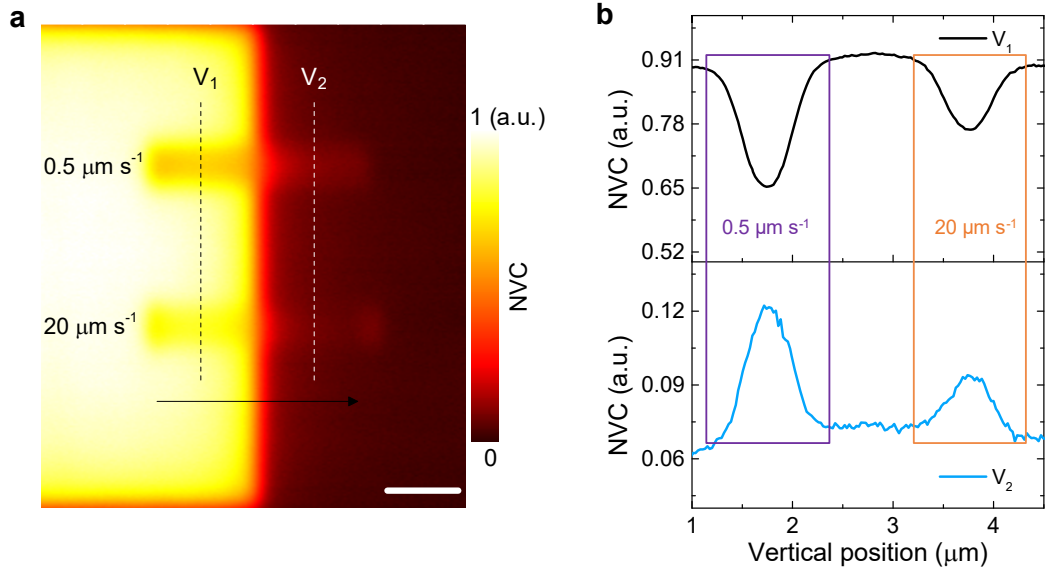

**Supplementary Figure 11 | The scanning velocity dependence of vacancy redistribution.**
**a,** The normalised vacancy concentration (NVC) map after mechanical scans were performed with two different scanning velocities (0.5 and 20 μm s$^{-1}$) across the boundary between a $V_O^{\cdot\cdot}$-enriched and pristine regions. We used the 120-uc thick STO film and a sharp tip for this experiment. Scans were performed over a 3x0.5 μm$^2$ area with a contact force of 6 μN. The force was applied only during the trace, and the tip was lift-off during the retrace. Fives lines were traced from left-right as marked by the horizontal arrow. **b**, NVC profiles along lines $V_1$ and $V_2$ showing a decrease (increase) in vacancy concentration inside (outside) the vacancy enriched region. The profiles are averaged over a 0.5 μm wide window. For clarity, we marked the regions scanned with velocities 0.5 and 20 μm s$^{-1}$ by purple and orange rectangles, respectively. Evidently, the net change in NVC inside and outside the vacancy enriched region is larger for the smaller scanning velocity. This implies the longer the tip stays in contact with the $V_O^{\cdot\cdot}$-enriched surface, a larger number of $V_O^{\cdot\cdot}$ migrate into the bulk from the contact point, and simultaneously an increased number of vacancies accumulate around the contact edge, yielding an enhanced lateral migration. Notably, unlike the experiments discussed in the main text (Figs. 2 and 4), where force was applied both during the race and retrace of the tip, in this controlled experiment the force was applied only during the trace. However, the NVC profiles in **b** clearly demonstrate the surface-bulk migration dominates over the lateral migration of vacancies with the tip. This observation further highlights the dominating role of the depolarisation field underneath a sharp tip. Note the region scanned with a tip velocity of 20 μm s$^{-1}$ shows a large increase in NVC at the end of the trace. This is an artefact, caused by a larger deformation of the STO surface during the withdrawal of the tip from the surface. We often made similar observations even for an engagement of the tip on the pristine region. For profiling the NVC map in (**a**), we have thus excluded the left and right ends of the scanned regions. The scale bar in **a** represents 1 μm.

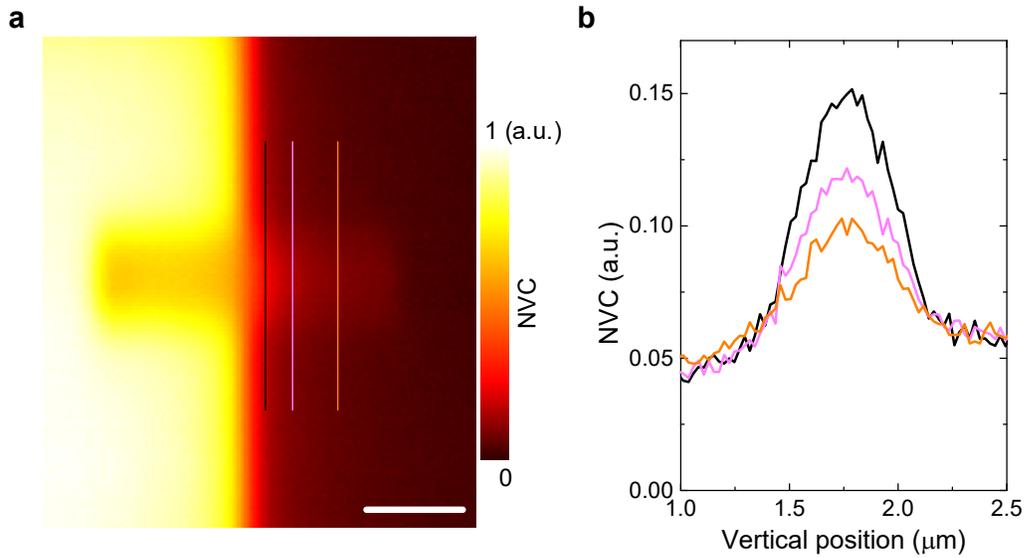

**Supplementary Figure 12 | Evolution of vacancy concentration with distance from the vacancy-enriched region. a,** Part of the normalised vacancy concentration (NVC) map shown in the supplementary figure 11. This image magnifies the region, which was scanned with a tip velocity of 0.5 µm s$^{-1}$. **b**, NVC profiles along lines marked on the NVC map in the black, red, and green colour, respectively. The lines are placed at a distance of about 0.2 µm, 0.5 µm, and 1 µm from the $V_o^{\cdot\cdot}$-enriched region. Evidently, the net gain in NVC systematically decreases with increasing distance from the $V_o^{\cdot\cdot}$-enriched region. We made a similar observation for the area scanned with the velocity of 20 µm s$^{-1}$. Note that the background NVC level in the pristine region uniformly decreases with increasing distance from the $V_o^{\cdot\cdot}$-enriched region. Thus, for this comparison we overlapped the background along the three lines in **a**. The scale bar in **a** represents 1 µm.

**Supplementary Discussion: Role of the applied tip bias during KPFM imaging**

Here, we address the influence of applied tip bias on KPFM images. We carried out this additional study to check whether the polarisation of STO film induced by the electric field from a biased tip affects the measured KPFM signal[18]. Note that for our work, we performed all KPFM measurements in the non-contact mode, whereby the tip and the STO surface were separated by an air-gap of 30 nm.

For an applied tip bias $V$ (= $V_{ac}$ + $V_{dc}$) and tip radius $R$, ignoring the radial distribution, the magnitude of the tip bias-induced electric field as a function of the tip-surface separation (lift height) $z$ can be approximated as $E(z) = \frac{2RV}{z^2}$ [19]. Therefore, by comparing KPFM scans performed at different lift heights, we can readily assess how much the tip bias affects the measurement. Following this logic, we took KPFM images across a $V_o^{..}$-enriched region of the 120-uc thick STO film by varying the lift height between 10-50 nm.

Supplementary Figs. 13 a-b show the representative KPFM image (taken at $z$ = 10 nm) and corresponding KPFM profiles. Evidently, the KPFM signal of the pristine STO surface uniformly changes by about 100 mV upon decreasing $z$ from 50 nm to 10 nm. This change is marginal, only 10% of the measured value (~ 1 V). Meanwhile, the KPFM contrast across the $V_o^{..}$-enriched region shows a relatively larger but uniform change by about 500 mV, which is about 30% of the KPFM contrast obtained at $z$ = 50 nm. Notably, during the KPFM imaging, the voltage ($V_{dc}$) that the feedback loop applies to the tip for nullifying the force between the tip and the STO surface corresponds to the measured KPFM signal. A simple algebraic calculation suggests that the strength of the tip bias-induced electric field should be 25 times larger at $z$ = 10 nm than at 50 nm: $E$ ($z$ = 10 nm) = 25$E$ ($z$ = 50 nm). Comparatively, however, the corresponding change in the measured KPFM signal (within the pristine region) and KPFM contrast (across the $V_o^{..}$-enriched region) is negligible.

Before concluding this discussion, let us briefly comment on the relatively larger effect of $z$-variation on the KPFM contrast. $V_o^{..}$ are electrically charged, and the electric field emanating from them would induce an electrostatic force between the tip and $V_o^{..}$. The magnitude of this force would depend on $z$, and on the concentration of $V_o^{..}$. The magnitude of this force, however, would be independent of the applied tip bias ($V$) [20]. Thus, during KPFM imaging, the closer the tip approaches towards the $V_o^{..}$-enriched region, the stronger it feels this $V_o^{..}$-induced force. Accordingly, the feedback loop utilizes a larger $V_{dc}$ for nullifying the force acting between the tip and the $V_o^{..}$-enriched STO surface. These considerations readily corroborate the relatively larger effect of $z$-variation on the KPFM contrast.

To sum up, based on our experiments we conclude that the influence of the applied tip bias and the resulting polarisation of STO is rather weak and, more importantly, is uniform across an entire KPFM image. While defining the concentration of $V_o^{..}$ or creating NVC maps from KPFM images, we defined the concentration of $V_o^{..}$ to be equal to the difference between the KPFM signals measured from the pristine and $V_o^{..}$-enriched regions. This approach readily removes this uniform contribution of the tip bias.

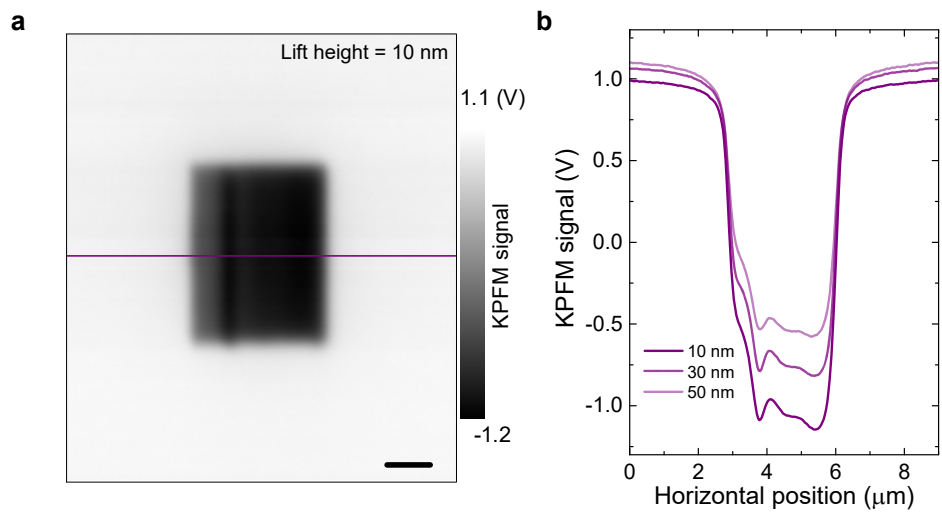

**Supplementary Figure 13 | KPFM imaging by varying the lift height a,** KPFM image of a $V_O^{\cdot\cdot}$-enriched STO surface. This image was obtained with a lift height of 10 nm. **b,** The KPFM profiles (along the line cut in **a**) from KPFM scans that were performed by varying the lift height between 10-50 nm. The scale bar in **a** represents 1 μm.

**Supplementary References**

1. Schie, M., Marchewka, A., Müller, T., De Souza, R. a & Waser, R. Molecular dynamics simulations of oxygen vacancy diffusion in SrTiO$_3$. *J. Phys. Condens. Matter* **24,** 485002 (2012).

2. Andrä, M. *et al.* The influence of the local oxygen vacancy concentration on the piezoresponse of strontium titanate thin films. *Nanoscale* **7,** 14351–14357 (2015).

3. Bieger, T., Maier, J. & Waser, R. Kinetics of oxygen incorporation in SrTiO$_3$ (Fe-doped): an optical investigation. *Sensors Actuators B Chem.* **7,** 763–768 (1992).

4. Merkle, R. & Maier, J. How is oxygen incorporated into oxides? A comprehensive kinetic study of a simple solid-state reaction with SrTiO$_3$ as a model material. *Angew. Chemie Int. Ed.* **47,** 3874–3894 (2008).

5. Lee, K. Y. *et al.* Controllable charge transfer by ferroelectric polarization mediated triboelectricity. *Adv. Funct. Mater.* **26,** 3067–3073 (2016).

6. Müller, K. A. & Burkard, H. SrTiO$_3$ : An intrinsic quantum paraelectric below 4 K. *Phys. Rev. B* **19,** 3593–3602 (1979).

7. Haeni, J. H. *et al.* Room-temperature ferroelectricity in strained SrTiO$_3$. *Nature* **430,** 758–761

(2004).

8. Kim, Y. S. *et al.* Localized electronic states induced by defects and possible origin of ferroelectricity in strontium titanate thin films. *Appl. Phys. Lett.* **94,** 202906 (2009).

9. Lee, D. *et al.* Emergence of room-temperature ferroelectricity at reduced dimensions. *Science* **349,** 1314–1317 (2015).

10. Lu, H. *et al.* Mechanical writing of ferroelectric polarization. *Science* **336,** 59–61 (2012).

11. Li, Y. L. *et al.* Phase transitions and domain structures in strained pseudocubic (100) $SrTiO_3$ thin films. *Phys. Rev. B* **73,** 184112 (2006).

12. Jesse, S., Baddorf, A. P. & Kalinin, S. V. Dynamic behaviour in piezoresponse force microscopy. *Nanotechnology* **17,** 1615–1628 (2006).

13. Balke, N. *et al.* Exploring local electrostatic effects with scanning probe microscopy: Implications for piezoresponse force microscopy and triboelectricity. *ACS Nano* **8,** 10229–10236 (2014).

14. Stengel, M. Surface control of flexoelectricity. *Phys. Rev. B* **90,** 201112 (2014).

15. Narvaez, J., Saremi, S., Hong, J., Stengel, M. & Catalan, G. Large flexoelectric anisotropy in paraelectric barium titanate. *Phys. Rev. Lett.* **115,** 37601 (2015).

16. Alexandrov, V. M. & Pozharskii, D. A. *Three-dimensional contact problems*. (Springer Netherlands, 2001).

17. Moos, R. & Hardtl, K. H. Defect chemistry of donor-doped and undoped strontium titanate ceramics between 1000° and 1400°C. *J. Am. Ceram. Soc.* **80,** 2549–2562 (1997).

18. Nielsen, D. A., Popok, V. N. & Pedersen, K. Modelling and experimental verification of tip-induced polarization in Kelvin probe force microscopy measurements on dielectric surfaces. *J. Appl. Phys.* **118,** 195301 (2015).

19. Gruverman, A., Auciello, O. & Tokumoto, H. Imaging and control of domain structures in ferroelectric thin films via scanning force microscopy. *Annu. Rev. Mater. Sci.* **28,** 101–123 (1998).

20. Orihuela, M. F., Somoza, A. M., Colchero, J., Ortuño, M. & Palacios-Lidón, E. Localized charge imaging with scanning Kelvin probe microscopy. *Nanotechnology* **28,** 25703 (2017).